\documentstyle[12pt,amsmath,amssymb,epic]{article}

\newcommand{\bbbone}{\mathchoice {\rm 1\mskip-4mu l} {\rm 1\mskip-4mu l}
{\rm 1\mskip-4.5mu l} {\rm 1\mskip-5mu l}}
\newcommand{\scalprod}[2]{\left\langle {#1}, {#2}\right\rangle}
\newcommand{\dom}{{\cal D}}
\newcommand{\RE}{{\rm Re}}
\newcommand{\IM}{{\rm Im}}
\newcommand{\fer}[1]{(\ref{#1})}
\newcommand{\ran}{{\rm Ran\,}}
\newcommand{\repsilonbar}{\,\overline{\!R}_\epsilon}
\newcommand{\repsilon}{R_\epsilon}

\newcommand{\h}{{\cal H}}
\newcommand{\cx}{{\mathbb C}}
\newcommand{\r}{{\mathbb R}}

\renewcommand{\d}{{\rm d}}

\newcommand{\Pbar}{\overline{P}}

\newcommand{\dirint}{\int^\oplus}

\newcommand{\av}[1]{\left\langle{#1}\right\rangle}

\newcommand{\mm}{{\frak M}}

\newcommand{\cc}{{\cal C}}
\newcommand{\ff}{{\cal F}}

\newcommand{\reta}{{\r_\eta}}
\newcommand{\psiann}{{\psi_{\alpha,\nu,n}}}
\newcommand{\tr}{{\rm tr\,}}

\newcommand{\rhocrit}{{\rho_{\rm crit}}}
\newcommand{\Omegaf}{{\Omega_{\cal F}}}
\newcommand{\1}{{\bf 1}}
\newcommand{\Wf}{{W_{\cal F}}}
\renewcommand{\L}{{\cal L}}
\newcommand{\varphif}{{\varphi_{\cal F}}}
\newcommand{\af}{{a_{\cal F}}}

\newcommand{\Oblcond}{\Omega_{\beta,\lambda}^{\rm con}}
\newcommand{\Llt}{L_{\lambda,\xi}}
\newcommand{\Obocond}{\Omega_{\beta,0}^{\rm con}}
\newcommand{\ocond}{\omega^{\rm con}}
\newcommand{\mbcond}{\mm_{\beta}^{\rm con}}

\renewcommand{\O}[1]{O\left({#1}\right)}

\newcounter{resultcounter} 
\stepcounter{resultcounter}

\newtheorem{theorem}{Theorem}[section]
\newtheorem{lemma}{Lemma}[section]

\begin{document}

\setcounter{page}{0}

\title{\vspace*{-2.5cm} Stability of Equilibria with a Condensate}

\author{ 
Marco Merkli \vspace*{.5cm} \footnote{Supported by a CRM-ISM postdoctoral fellowship and by McGill University; merkli@math.mcgill.ca; http://www.math.mcgill.ca/$\sim$merkli/}\\
Dept. of Mathematics and Statistics, McGill University\\
805 Sherbrooke  W., Montreal\\
Canada, QC, H3A 2K6\\
\vspace{-.3cm}
\centerline{and}
\and
Centre de Recherches Math\'ematiques, 
Universit\'e de Montr\'eal\\
Succursale centre-ville, Montr\'eal\\
Canada, QC, H3C 3J7
}

\maketitle
\vspace{-1cm}
\begin{abstract}
We consider a quantum system composed of a spatially infinitely extended free Bose gas with a condensate, interacting with a small system (quantum dot) which can trap finitely many Bosons. Due to spontaneous symmetry breaking in the presence of the condensate, the system has many equilibrium states for each fixed temperature. 
We extend the notion of Return to Equilibrium to systems possessing a multitude of equilibrium states and show in particular that a condensate coupled to a quantum dot has the property of Return to Equilibrium in a weak coupling sense: any local perturbation of an equilibrium state of the coupled system, evolving under the interacting dynamics, converges in the long time limit to an asymptotic state. The latter is, modulo an error term, an equilibrium state which {\it depends} in an explicit way on the local perturbation (an effect due to long-range correlations). The error term vanishes in the small coupling limit.\\
\indent
We deduce the stability result from properties of structure and regularity of eigenvectors of the generator of the dynamics, called the Liouville operator.    
Among our technical results is a Virial Theorem for Liouville type operators which has new applications to systems with and without a condensate.
\end{abstract}
\thispagestyle{empty}
\setcounter{page}{1}
\setcounter{section}{1}


\setcounter{section}{0}

\section{Introduction}

This paper is devoted to the study of the dynamics of a class of quantum systems consisting of a small part in interaction with a large heat reservoir, modelled by an infinitely extended ideal gas of Bosons. We further develop spectral methods in the framework of algebraic quantum field theory and apply them to the class of systems at hand, for which the already existing techniques cannot be applied. \\
\indent
Our main physical interest is the long-time behaviour of initial states close to an equilibrium state of a Bose gas that is so dense (for fixed temperature) or so cold (for fixed density) that it has a Bose-Einstein condensate, in interaction with a small system that can store a finite number of Bosons. The condensate induces {\it long-range correlations} in the system and as a consequence there are many equilibrium states for a fixed value of the temperature $T=1/\beta$. It is thus necessary to extend the notion of {\it Return to Equilibrium} to systems possessing a multitude of equilibrium states. In general a system has a special class of equilibrium states (extremal ones) whose superpositions generate all equilibrium states. It is reasonable to expect (and proven for the model considered here) that each extremal equilibrium state has the property of Return to Equilibrium. This leads to a general definition of this property which we introduce in Section \ref{rtesec}. A feature of this situation is that starting with a local perturbation of a given superposition of extremal equilibrium states the system converges in the long time limit to a possibly {\it different} superposition of extremal equilibrium states (redistribution of phases). The asymptotic state depends thus on the initial condition. This limitation of the dispersive nature of a system is natural in view of the long-range correlations. \\
\indent
One of our goals is to prove {\it weak coupling return to equilibrium}, in the above setting,  saying that any initial condition close to an equilibrium state of the coupled system, evolving under the coupled dynamics, converges in the long time limit to an asymptotic state. The latter is again an equilibrium state (different from the initial one), modulo an error which disappears in the limit of small coupling.  We expect a stronger result to hold, namely that any initial condition close to the interacting system converges, in the long time limit, to an equilibrium state of the interacting system. This result  has been obtained for systems without a condensate in a variety of recent papers, [JP1,BFS,M1,DJ,FM2].
It is surprising that none of the methods developed in these references -- nor elsewhere, according to our knowledge --  can be applied to the present case. This is due to the fact that the {\it form factor} of the interaction, a coupling function $g\in L^2(\r^3,d^3k)$, whose properties are dictated by physics, exhibits the infrared behaviour $0<|g(0)|<\infty$. It lies in between the two ``extreme'' behaviours $g(0)=0$ (more precisely, $g(k)\sim |k|^p$, some $p>0$, as $|k|\sim 0$) and $|g(0)|=\infty$ (more precisely, $g(k)\sim |k|^{-1/2}$ as $|k|\sim 0$), which are the only ones that can be treated using approaches existing so far. We give here a partial remedy to this situation by establishing a ``positive commutator theory'' (a first step in a Mourre theory) which is applicable to a wide variety of interactions, including the case where $g(0)$ is a nonzero, finite constant. Our remedy is only partial in that so far, we show that the equilibrium state is stable, in the sense mentioned above, but we cannot prove return to equilibrium. The obstruction seems to be of technical nature (see Section \ref{subsubdiscussion} for a discussion of this point).\\
\indent
After clarifying (defining) the notion of Return to Equilibrium in the setting of multiple KMS states our analysis consists of two main steps. The first one is to carry out the decomposition of an arbitrary equilibrium state into extremal ones and to describe the dynamics of each of them. The second step in our analysis, which is the main part of this paper, consists in analyzing the time asymptotic behaviour of each extremal equilibrium state. We do this by examining the spectrum of the Liouville operators generating the dynamics. Our approach gives an extension of the positive commutator method, including a new virial theorem which yields improved results in the analysis of related problems for systems without a condensate.

\subsection{An extended notion of Return to Equilibrium}

We review the notion of {\it Return to Equilibrium} and extend it to systems having multiple KMS states. Our guiding example is a reservoir of free Bosons where the non-uniqueness of KMS states is due to spontaneous gauge-group symmetry breaking in the presence of a Bose-Einstein Condensate.

\subsubsection{Free Bose gas and Bose-Einstein Condensation}
\label{fbg}

The kinematical algebra describing the Bose gas is the {\it Weyl algebra} ${\frak W}(\dom)$ over a suitably chosen test-function space of one-particle wave functions $\dom \subset L^2(\r^3,d^3k)$. ${\frak W}(\dom)$ is generated by Weyl operators $W(f)$, $f\in\dom$, satisfying the canonical commutation relations (CCR)
\begin{equation}
W(f)W(g)=e^{-\frac{i}{2}\IM\scalprod{f}{g}} W(f+g),
\label{106}
\end{equation}
where $\scalprod{\cdot}{\cdot}$ is the inner product induced by $L^2(\r^3,d^3k)$. The dynamics of the Bose gas is given by the Bogoliubov transformation 
\begin{equation}
W(f)\mapsto \alpha_t(W(f))=W(e^{it\omega}f),
\label{bogdyn}
\end{equation}
where 
\begin{equation}
\omega(k)= |k|^2,\ \ \ \mbox{or}\ \ \ \omega(k)=|k|.
\label{109}
\end{equation}
The first choice in \fer{109} describes non-relativistic Bosons, while the second one describes massless relativistic ones.

We outline first the construction of the equilibrium state of [AW], which gives a good physical understanding of the emergence of a condensate, and then we relate this to the works of [C] and [LP]. 
Any state $\omega$ on the Weyl algebra ${\frak W}(\dom)$ is uniquely determined by its so-called {\it generating} (or expectation) {\it functional} $E:\dom\rightarrow\cx$, given by 
\begin{equation}
\omega(W(f))= E(f),
\label{107}
\end{equation}
and conversely, if $E:\dom\rightarrow\cx$ is a (non linear) function satisfying certain compatibility conditions then it defines uniquely a state on ${\frak W}(\dom)$, see e.g. [A, M2]. \\
\indent
Let $\r^3 \ni k\mapsto \rho(k)>0$ be a given function (the ``continuous momentum-density distribution''), and $\rho_0\geq 0$ a fixed number (the ``condensate density''). Araki and Woods [AW] obtain a state of the Bose gas by the following procedure. Restrict the gas to a periodic box $\Lambda$ of volume $V$ in $\r^3$ and put $V\rho_0$ particles in the ground state of the one particle Hamiltonian $H_\Lambda=-\Delta$ (or $H_\Lambda=\sqrt{-\Delta}$), and a discrete distribution of particles in excited states. Then take the limit $V\rightarrow\infty$ while keeping $\rho_0$ fixed and letting the discrete distribution of excited states tend to $\rho(k)$. Like this [AW] obtain the family of generating functionals 
\begin{equation}
E^{\rm AW}_{\rho,\rho_0}(f)= \exp\left[ -\frac{1}{4} \scalprod{f}{\left(1+2(2\pi)^3\rho\right)f}\right]J_0\left(\sqrt{2(2\pi)^3\rho_0}|f(0)|\right), 
\label{aw}
\end{equation}
where $J_0(\sqrt{\alpha^2+\beta^2})=\int_{-\pi}^\pi\frac{d\theta}{2\pi} e^{-i(\alpha\cos\theta +\beta\sin\theta)}$, $\alpha,\beta\in\r$  (Bessel function). 
$E^{\rm AW}_{\rho,\rho_0}$ defines uniquely a state of the infinitely extended Bose gas according to \fer{107}. The physical interpretation is that the resulting state describes a free Bose gas where a sea of particles, all being in the same state (corresponding to the ground state of the finite-volume Hamiltonian), form a condensate with density $\rho_0$, which is immersed in a gas of particles where $\rho(k)$ particles per unit volume have momentum in the infinitesimal volume $d^3k$ around $k\in\r^3$. Since the Hamiltonian in the finite box is taken with periodic boundary conditions the condensate is homogeneous in space (the ground state wave function is a constant in position space). The resulting state is an equilibrium state (satisfies the KMS condition) if the momentum density distribution is given by 
\begin{equation}
\rho(k)=(2\pi)^{-3}\frac{1}{e^{\beta\omega(k)}-1},
\label{115}
\end{equation}
corresponding to Planck's law of black body radiation. 
\\
\indent
According to the principles of quantum statistical mechanics, the equilibrium state of the infinite system is obtained by taking the thermodynamic limit of local canonical or grand-canonical Gibbs states. This has been done in [C] (canonical) and [LP] (grand-canonical). We first outline the result of [LP]. The density matrix (acting on Fock space) for the local system is 
\begin{equation}
\sigma_{\beta,z}^\Lambda = \frac{e^{-\beta(H_\Lambda-\mu N_\Lambda)}}{\tr e^{-\beta(H_\Lambda-\mu N_\Lambda)}},
\label{r5}
\end{equation}
where $z=e^{\beta\mu}$ is the fugacity and $N_\Lambda$ is the number operator. For a fixed inverse temperature $0<\beta<\infty$ define the {\it critical density} by 
\begin{equation}
\rhocrit(\beta)=(2\pi)^{-3}\int \frac{d^3k}{e^{\beta\omega}-1},
\label{112}
\end{equation}
and denote by $\overline\rho$ the total (mean) density of the gas. Then 
\begin{equation}
E_{\beta,z}^\Lambda(f):=\tr\left( \sigma_{\beta,z}^\Lambda W(f)\right) \longrightarrow E_{\beta,\overline\rho}(f),
\label{f6}
\end{equation}
where the limit $|\Lambda|\rightarrow\infty$ is taken with the constraint  $\overline\rho = \frac{\tr(\sigma_{\beta,z(\Lambda)}^\Lambda N_\Lambda)}{\tr(\sigma_{\beta,z(\Lambda)}^\Lambda)}$, determining the value of $z(\Lambda)$. The limiting generating functional \fer{f6} is 
\begin{equation}
E_{\beta,\overline\rho}(f) = 
\left\{
\begin{array}{cl}
e^{-\frac{1}{4}\|f\|^2} e^{-\frac{1}{2}\scalprod{f}{\frac{z_\infty}{e^{\beta\omega}-z_\infty}f}}, & \overline\rho\leq\rhocrit(\beta)\\
E^{\rm con}_{\beta,\overline\rho}(f), & \overline\rho\geq \rhocrit(\beta)
\end{array}
\right.
\label{r7}
\end{equation}
where, with $\rho_0=\overline\rho-\rhocrit(\beta)\geq 0$, 
\begin{equation}
E^{\rm con}_{\beta,\overline\rho}(f) = \exp\left\{-\frac{1}{4}\|f\|^2\right\} \exp\left\{-\frac{1}{2}\scalprod{f}{(2\pi)^3\rho f}\right\} \exp\left\{-4\pi^3\rho_0|f(0)|^2\right\},
\label{r8}
\end{equation}
and  $\rho=\rho(k)$ is given in \fer{115}. 
For subcritical density, the number $z_\infty\in [0,1]$ is determined by the equation
\begin{equation}
\overline\rho =(2\pi)^{-3}\int\frac{z_\infty}{e^{\beta\omega}-z_\infty}d^3k.
\label{114}
\end{equation}
In the supercritical case we have $z_\infty=1$ which corresponds to a vanishing chemical potential, $\mu_\infty=0$.\\
\indent
The thermodynamic limit of the canonical local Gibbs state is treated in [C], the density matrix is 
\begin{equation}
\mu_{\beta,\rho}^\Lambda=\frac{ e^{-\beta H_\Lambda}P_{\rho V}}{\tr  e^{-\beta H_\Lambda}P_{\rho V}},
\label{116}
\end{equation}
and  
$P_{\rho V}$ is the projection onto the subspace of Fock space with $\rho V$ particles (if $\rho V$ is not an integer take a convex combination of canonical states with integer values $\rho_1V$ and $\rho_2 V$ extrapolating $\rho V$). 
The limiting generating functional is given by 
\begin{equation}
E^{\rm C}_{\beta,\rho}(f)=
\left\{
\begin{array}{cl}
e^{-\frac{1}{4}\|f\|^2} e^{-\frac{1}{2}\scalprod{f}{\frac{z_\infty}{e^{\beta\omega}-z_\infty}f}}, & \rho\leq\rhocrit(\beta)\\
E^{\rm AW}_{\rho,\rho_0}(f), & \rho\geq \rhocrit(\beta).
\end{array}
\right.
\label{117}
\end{equation}
It coincides with the grand-canonical generating functional in the subcritical case, and with the Araki-Woods generating functional with $\rho$ given by \fer{115} and 
\begin{equation*}
\rho_0=\rho-\rhocrit
\end{equation*}
in the supercritical case. To complete the exposition of this triangle of generating functionals we mention that (see [C]) the grand-canonical and the canonical generating functionals are linked in the supercritical case, $\rho_0>0$, by the Laplace transform
\begin{equation}
E_{\beta,\overline\rho}^{\rm con}(f)=\int_0^\infty K(r;\overline\rho) E_{\beta,r}^{\rm C}(f) dr,
\label{r9}
\end{equation}
where the {\it Kac density} $K(r;\overline\rho)$ is
\begin{equation}
K(r;\overline\rho)=
\left\{
\begin{array}{cl}
e^{-(r-\rhocrit)/\rho_0}/\rho_0, & r>\rhocrit\\
0 & r\leq\rhocrit
\end{array}
\right.
\label{r10}
\end{equation}
This means that the grand-canonical equilibrium state with supercritical mean density $\overline\rho$ is a superposition of canonical equilibrium states with supercritical densities $r$, weighted with the Kac density $K(r,\overline\rho$). \\
\indent
To ease the notation we simply write $\omega_\beta$ for the equilibrium state corresponding to \fer{r9} (imagining a supercritical mean density $\overline\rho$ to be fixed). 
Space translations are given by $\tau_x f(y)=f(y-x)$, for $x\in\r^3$ ($\tau_x f(k)=e^{ikx}f(k)$ in Fourier space) and since $E^{\rm con}_{\beta,\overline\rho}(\tau_xf)=E^{\rm con}_{\beta,\overline\rho}(f)$ for all $x\in\r^3$ (see \fer{r8}) the state $\omega_\beta$ is space translation invariant. However, due to the presence of the condensate, the system has {\it long range correlations}:
\begin{eqnarray}
\lefteqn{
\lim_{|x|\rightarrow\infty} \omega_\beta\big(W(f)W(\tau_xg)\big)}\nonumber\\
&& = \omega_\beta\big(W(f)\big)\  \omega_\beta\big(W(g)\big)\ \exp\left[ -8\pi^3\rho_0 \RE\left( \overline f(0)g(0)\right)\right],
\label{r20.1'}
\end{eqnarray}
where $\rho_0=\overline\rho-\rhocrit$ is the condensate density. This means that $\omega_\beta$ is not a factorial state, i.e. the von Neumann algebra of observables represented in the Hilbert space associated to $({\frak W}(\dom), \omega_\beta)$ is not a factor (see for instance [Ha], Theorem 3.2.2). In Section \ref{mainresultsection} we decompose $\omega_\beta$ into a superposition of extremal factorial states $\omega_\beta^\xi$,
\begin{equation}
\omega_\beta(A)=\int_{\r^2}d\mu_{\beta,\overline\rho}(\xi) \omega_\beta^\xi(A),
\label{r20'}
\end{equation}
where the probability measure 
\begin{equation}
d\mu_{\beta,\overline\rho}(\xi) := K(r,\overline\rho)dr\frac{d\theta}{2\pi}
\label{mubetarho}
\end{equation}
is  supported on $\{\xi=(r,\theta)\in[\rhocrit,\infty)\times S^1\}\subset\r^2$. Each $\omega_\beta^\xi$ is a $\beta$-KMS state w.r.t. the dynamics \fer{bogdyn}, having the cluster property (compare with \fer{r20.1'}!) 
\begin{equation}
\lim_{|x|\rightarrow\infty}\omega_\beta^\xi\big(W(f)W(\tau_xg)\big)= \omega_\beta^\xi\big(W(f)\big)\ \omega_\beta^\xi\big(W(g)\big),
\label{cluster}
\end{equation}
which is 
also called the property of {\it strong mixing} w.r.t. space translations. \\
\indent
Clearly, the gauge transformations $\gamma_s(W(f))=W(e^{is}f)$, $s\in\r$, commute with the dynamics given in \fer{bogdyn}, $\alpha_t\circ\gamma_s=\gamma_s\circ\alpha_t$ for all $s,t\in\r$, and the state $\omega_\beta$ is invariant under $\gamma_s$. However, the equilibrium states $\omega_\beta^\xi$ are {\it not} invariant under $\gamma_s$; hence there is a family of equilibrium states $\omega_\beta^\xi$ possessing  ``less symmetry than the dynamics'', a property of the system called {\it spontaneous symmetry breaking}.\\
\indent
We can take {\it any} probability measure $\mu$, supported inside $[\rhocrit,\infty)\times S^1$, and define the $\beta$-KMS state
\begin{equation}
\omega^\mu(A) = \int_{\r^2}d\mu(\xi)\omega_\beta^\xi(A).
\label{r21'}
\end{equation}
This shows that there is, for each $\beta$ fixed, a multitude of equilibrium states (each determined by a $\mu$), and we must examine what this means for the notion of Return to Equilibrium.\\
\indent
{\it Remark.\ } The integration over $r>\rhocrit$ in \fer{r20'} comes from the use of the grand-canonical ensemble, see \fer{r9}. In the canonical case the value of $r$ is fixed and the integration is only over $\theta\in S^1$. Each state $\omega_\xi$ is extremal invariant for space translations $\tau_x$ ($\omega_\xi$ is $\tau_x$-invariant and cannot be decomposed into a convex combination of two other $\tau_x$-invariant states; this follows from the cluster property \fer{cluster}). As a consequence the average $\lim_{|\Lambda|\rightarrow\infty} \frac{1}{|\Lambda|}\int_\Lambda d^3x\, \pi_\xi(\tau_x(A))$ exists and is a multiple of the identity ($A$ a local observable). In particluar, for each $\xi$ fixed, the density $\rho$ and  the phase $\theta$ are observables in $\pi_\xi({\frak A})''$ which reduce to numbers. ($\rho$ is obtained by choosing $\tau_x(A)=n(x)=a^*(x)a(x)$ representing the number of particles at the point $x$, $e^{i\theta}$ is obtained for $\tau_x(A)=a(x)$, see [H].)
The density and the phase are observables which commute with all other observables (they belong to the center of the observable algebra associated with $\omega_\beta$) and are thus regarded as classical quantities. For instance, a maximal lack of knowledge of the value of $\theta$ is expressed by taking a superposition of $\omega_\xi$ where $\theta$ is uniformly distributed. In the state $\omega_\beta^\xi$ the values of density and phase are fixed.

\subsubsection{Return to Equilibrium}
\label{rtesec}

Consider first the equilibrium state $\omega_\beta$ describing a reservoir of free Bosons at inverse temperature $\beta$ with a fixed {\it subcritical density} $\overline\rho<\rhocrit$ (no condensate). This system has the property of Return to Equilibrium: for any observables $A,B$ we have 
\begin{equation}
\lim_{t\rightarrow\infty}\omega_\beta(B^*\alpha_t(A)B)=\omega_\beta(B^*B)\omega_\beta(A).
\label{a1}
\end{equation}
The functional $A\mapsto \omega_\beta(B^*AB)$ defines a vector state (if $\omega_\beta(B^*B)=1$) which is a local perturbation of $\omega_\beta$.  Equation \fer{a1} extends to the closure of all convex combinations of such vector states (called the set of $\omega_\beta$-normal states). 
Property \fer{a1} has been shown to hold also for reservoirs (without condensates) {\it interacting with a small system}; then $\omega_\beta$ in \fer{a1} is the equilibrium state of the interacting system and $\alpha_t$ is the coupled dynamics, see [JP1,BFS,DJ,M,FM2]. For different (scattering) approaches to similar problems we refer to [R,LV], and to [HL,QV] for stochastic methods. \\
\indent
Consider next the Bose gas is in a state $\omega_\beta$ {\it with} a condensate. As we have seen above there are many $\beta$-KMS states w.r.t. the free dynamics (and $\beta$ fixed), each characterized by a probability measure $\mu$ as in \fer{r21'}. 
Fix such a state $\omega^\mu$. 
It is easy to see that each of the extremal equilibrium states $\omega_\beta^\xi$ has the property of return to equilibrium, i.e., $\lim_{t\rightarrow\infty}\omega_\beta^\xi(B^*\alpha_t(A)B)=\omega_\beta^\xi(B^*B)\omega_\xi(A)$, for all $\xi,A,B$. It follows that 
\begin{equation}
\lim_{t\rightarrow\infty}\omega^\mu(B^*\alpha_t(A)B)=\int_{\r^2}d\mu(\xi)
\ \omega_\beta^\xi(B^*B) \,\omega_\beta^\xi(A).
\label{a3}
\end{equation}
The r.h.s. of \fer{a3} is in general not equal to $\omega^\mu(A)$: the limiting state depends on the initial condition (i.e., on $\omega_\beta^\xi(B^*B)$). Relation \fer{a3} motivates the following abstract \\

{\bf Definitions.\ }
 {\bf 1.} Let $\omega$ be a state on a $C^*$algebra $\frak A$, invariant w.r.t. a $*$automorphism group $\alpha_t$ of $\frak A$. We say that {\it $\omega$ is asymptotically stable (w.r.t. $\alpha_t$)} if $\lim_{t\rightarrow\infty}\omega(B^*\alpha_t(A)B)=\omega(B^*B)\omega(A)$, for any $A, B\in\frak A$.\\
\indent
{\bf 2.} 
Let $\omega_\xi$, $\xi\in X$ (a measurable space), be a measurable family of states on a $C^*$algebra $\frak A$ (in the sense that $\xi\mapsto \omega_\xi(A)$ is measurable for all $A\in\frak A$) and let $\alpha_t$ be a $*$automorphism group of $\frak A$. Given any probability measure $\mu$ on $X$ we define the state  
\begin{equation}
\omega^\mu = \int_X d\mu(\xi)\ \omega_\xi.
\label{a4}
\end{equation}
We say that {\it the family $\omega_\xi$ is asymptotically stable (w.r.t. $\alpha_t$)} if, for any $\mu, A, B$, we have 
\begin{equation}
\lim_{t\rightarrow\infty}\omega^\mu(B^*\alpha_t(A)B)=\int_X d\mu(\xi)\ \omega_\xi(B^*B)\ \omega_\xi(A).
\label{a5}
\end{equation}
{}\indent
{\bf 3.} If $\omega$ in 1. is a $(\beta,\alpha_t)$-KMS state then we say {\it $\omega$ has the poperty of Return to Equilibrium}. Similarly, if the $\omega_\xi$ in 2. are $(\beta,\alpha_t)$-KMS states (then so is $\omega^\mu$) we say {\it the family $\omega_\xi$ has the property of Return to Equilibrium}.\\

{\it Remarks.\ } {\it 1.} More generally one could consider in \fer{a4} the case where $\mu$ is a measure on the space of all states on $\frak A$. The present setup is sufficient for our purposes. \\
\indent
{\it 2.} If $\omega_\xi$ is asymptotically stable w.r.t. $\alpha_t$, for all $\xi$, then the family $\omega_\xi$ is asymptotically stable.\\
\indent
{\it 3.} If $B$ satisfies $\omega_\xi(B^*B)=1$ for all $\xi$ (say if $B$ is unitary) then \fer{a5} means that $\omega^\mu$ is asymptotically stable. In general, the effect of the initial condition on the limit state \fer{a5} is a redistribution of the relative weights.\\

In the above definitions the dynamics of the system is given by a (not necessarily norm continuous) $*$automorphism group $\alpha_t$ of a $C^*$algebra $\frak A$. While this description applies to free Fermionic or Bosonic heat reservoirs it does not in our case of interest, where a {\it Bosonic} reservoir is {\it coupled} to a small system. The problem is that one does not know how to define the dynamics for the coupled system as a $*$automorphism group of the $C^*$algebra of observables (unless the algebra is changed, see [FM1]). One circumvents this issue by defining the interacting dynamics, via a converging perturbation series, as a $*$automorphism group of the {\it von Neumann algebra} associated with a reference state (the uncoupled equilibrium state), see Section \ref{intsyst}. 
We shall therefore adapt the above definitions to a setting where the dynamics is not defined on the level of the $C^*$algebra of observables, but is rather expressed as a (``Schr\"odinger'') dynamics of states.\\

{\bf Definitions.\ }
 {\bf 1'.} Let $\omega$ be a state on a $C^*$algebra $\frak A$ and denote by $({\cal H}_\omega, \pi_\omega,\Omega_\omega)$ its GNS representation, $\omega(A)=\scalprod{\Omega_\omega}{\pi_\omega(A)\Omega_\omega}$. Suppose $\sigma_t$ is a $*$automorphism group of the von Neumann algebra $\pi_\omega({\frak A})''$. We say that {\it $\omega$ is asymptotically stable (w.r.t. $\sigma_t$)} if 
\begin{equation}
\lim_{t\rightarrow\infty}\scalprod{\Omega_\omega}{\pi_\omega(B^*)\sigma_t(\pi_\omega(A))\pi_\omega(B)\Omega_\omega}=\omega(B^*B)\omega(A),
\label{a6}
\end{equation}
for all $A,B\in\frak A$.\\
\indent
{\bf 2'.} 
Let $\omega_\xi$, $\xi\in X$ (a measurable space), be a measurable family of states on a $C^*$algebra $\frak A$ and denote their GNS representations by $({\cal H}_\xi,\pi_\xi,\Omega_\xi)$. Suppose that, for each $\xi$, $\sigma_t^\xi$  is a $*$automorphism group of the von Neumann algebra $\pi_\xi({\frak A})''$.
We say that {\it the family $\omega_\xi$ is asymptotically stable (w.r.t. $\sigma^\xi_t$)} if, for any $\mu, A, B\in\frak A$, we have 
\begin{equation}
\lim_{t\rightarrow\infty}\int_X d\mu(\xi) \scalprod{\Omega_\xi}{\pi_\xi(B^*) \sigma^\xi_t(\pi_\xi(A)) \pi_\xi(B) \Omega_\xi} =
\int_X d\mu(\xi)\ \omega_\xi(B^*B)\ \omega_\xi(A),
\label{a7}
\end{equation}
where $\mu$ is an arbitrary probability measure on $X$.\\
\indent
{\bf 3'.} If $\omega$ in 1'. is a $(\beta,\sigma_t)$-KMS state of $\pi_\omega({\frak A})''$ then we say {\it $\omega$ has the poperty of Return to Equilibrium}. Similarly, if the $\omega_\xi$ in 2'. are $(\beta,\sigma^\xi_t)$-KMS states of $\pi_\xi({\frak A})''$ we say {\it the family $\omega_\xi$ has the property of Return to Equilibrium}.\\

{\it Remark.\ } In case $\sigma_t^\xi(\pi_\xi(A))=\pi_\xi(\alpha_t(A))$ for some $*$automorphism group $\alpha_t$ of $\frak A$, the second set of definitions reduces to the first one.\\

In this paper we show a (weak) version of relation \fer{a7}, for equilibrium states of the Bose gas with a condensate interacting with a small system, where the time limit is taken in the ergodic mean sense and is followed by the limit of small coupling constant, see \fer{intro146}, \fer{intro146'}.

\subsection{A condensate coupled to a quantum dot (quantum tweezers) and its weak coupling stability}

In this section we outline the model and the stability result. For technical detail we refer the reader to Section \ref{mainresultsection}. The small system with which the supercritical Bose gas interacts can trap finitely many Bosons - we call it therefore a quantum dot. One can imagine the use of such a trap to remove single (uncharged) particles from a reservoir, hence the name quantum tweezers (see e.g. [DWRN]).\\
\indent
The pure states of the small system are given by normalized vectors in $\cx^d$. We interpret $[1,0,\ldots,0]$ as the {\it ground state} (or ``vacuum state''), $[0,1,0,\ldots,0]$ as the first excited state, e.t.c. The Hamiltonian is given by the diagonal matrix
\begin{equation}
H_1={\rm diag}(0,1,2,\dots,d-1).
\label{r1}
\end{equation}
Our method applies to any selfadjoint diagonal matrix with non-degenerate spectrum. We introduce the raising and lowering operators, $G_+$ and $G_-$,  
\begin{equation}
G_+=
\left[
\begin{array}{cccc}
0 & 0 & \cdots & 0\\
1 & 0 & \ddots & \vdots \\
\vdots& \ddots & \ddots & 0\\
0 & \cdots &1 & 0
\end{array}
\right],\ \ \ 
G_-=(G_+)^*,
\label{r3}
\end{equation}
($G_+$ has ones on its subdiagonal) 
which satisfy $H_1G_\pm = G_\pm(H_1\pm 1)$. The action of $G_+$ ($G_-$) increases (decreases) the excitation level by one. The dynamics of an observable $A\in{\cal B}(\cx^d)$ (bounded operators on $\cx^d$) is given by $A\mapsto \alpha_1^t(A)=e^{itH_1}Ae^{-itH_1}$.\\
\indent
 The observable algebra of the combined system is the $C^*$-algebra
\begin{equation}
{\frak A}={\cal B}(\cx^d)\otimes{\frak W}(\dom),
\label{123}
\end{equation}
and the non-interacting dynamics is the $*$automorphism group of $\frak A$ given by 
\begin{equation}
\alpha^t_0=\alpha^t_1\otimes\alpha^t_2,
\label{124}
\end{equation}
where we now denote the free field dynamics \fer{bogdyn} by $\alpha^t_2$. \\
\indent
Let $\mu$ be a fixed probability measure on $[\rhocrit,\infty)\times S^1$ and consider the $(\beta,\alpha_0^t)$-KMS state
\begin{equation}
\omega^{\rm con}_{\beta,0}=\int_{\r^2} d\mu(\xi) \ \omega_{1,\beta}\otimes\omega_{\xi},
\label{intro122}
\end{equation}
where $\omega_{1,\beta}$ is the $(\beta,\alpha^t_1)$-KMS state (Gibbs state) of the small system, and $\omega_\xi$ is a $(\beta,\alpha^t_2)$-KMS state with fixed density and phase (see Subsection \ref{fbg}). The subindex ``$0$'' in \fer{intro122} indicates the absence of an interaction. 
Let $\cal H$ denote the (GNS-) Hilbert space of state vectors obtained from the algebra $\frak A$, \fer{123}, and the state \fer{intro122}. Furthermore, let $\Obocond\in\h$ denote the cyclic vector in $\h$ representing the state $\ocond_{\beta,0}$, and let $\pi$ be the GNS representation of $\frak A$ on $\h$. Since $\ocond_{\beta,0}$ is invariant under $\alpha_0^t$ (see \fer{124}) there is a selfadjoint operator ${\cal L}_0$ acting on $\h$, called the thermal Hamiltonian or {\it Liouvillian}, satisfying
\begin{eqnarray}
\pi(\alpha_0^t(A)) &=&e^{it\L_0}\pi(A)e^{-it\L_0}, 
\label{intro126}\\
{\cal L}_0\Obocond &=& 0,
\label{intro127}\end{eqnarray}
for all $A\in\frak A$. 
In order to describe interactions between the small system and the Bose gas one replaces the (non-interacting) Liouvillian ${\cal L}_0$ by the (interacting) Liouvillian ${\cal L}_\lambda$, which is the selfadjoint operator on $\h$ given by 
\begin{equation}
{\cal L}_\lambda ={\cal L}_0+\lambda {\cal I},
\label{intro128}
\end{equation}
where $\lambda\in\r$ is a coupling constant and ${\cal I}$ is the operator on $\cal H$ determined by the formal interaction term
\begin{equation}
\lambda \big( G_+\otimes a(g) + G_-\otimes a^*(g)\big),
\label{144}
\end{equation}
where $G_\pm$ are the raising and lowering operators, \fer{r3}, and $a^\#(g)$ are creation ($\#=*$) and annihilation operators of the heat bath, smeared out with a function $g\in\dom$, called a {\it form factor}. The operator $G_+\otimes a(g)$ destroys a Boson and traps it in the quantum dot (whose excitation level is increased by one) and similarly, the effect of $G_-\otimes a^*(g)$ is to release a Boson from the quantum dot. The total number of particles, measured by the observable $H_1+\int_{\r^3} a^*(k)a(k)d^3k$, is preserved by the interaction (meaning that \fer{144} commutes with this operator). Since the quantum dot can absorb only finitely many Bosons, the interacting equilibrium state is a (local) perturbation of the non-interacting one. A physically different situation occurs when the condensate is coupled to another reservoir. We would then expect that time-asymptotic states are of {\it non-equilibrium stationary} nature, a situation we investigate in a different work. \\
\indent
Of course, \fer{144} has a meaning only in a regular representation of the Weyl algebra, e.g. the representation $\pi$ above, see Subsection \ref{intsyst}. The interaction ${\cal I}$ has the property that the dynamics generated by ${\cal L}_\lambda$ defines a $*$automorphism group $\sigma^t_\lambda$ of the von Neumann algebra $\mbcond\subset{\cal B}(\h)$ obtained by taking the weak closure of the algebra $\pi({\frak A})$. One can show that there exists a vector $\Oblcond\in\h$ defining a $(\beta,\sigma^t_\lambda)$-KMS state on $\mbcond$. We call $\Oblcond$ the perturbed KMS state, it satisfies 
\begin{equation}
{\cal L}_\lambda \Oblcond =0.
\label{intro145}
\end{equation}
As can be seen from \fer{intro122} the Hilbert space $\h$ has a direct integral decomposition,
\begin{equation}
\h=\dirint_{\r^2} d\mu(\xi)\ \h_\xi,
\label{intro200}
\end{equation}
where $\h_\xi$ is the Hilbert space associated with the state $\omega_{1,\beta}\otimes\omega_\xi$. Similarly, the representation $\pi$, the cyclic vectors $\Omega_{\beta,0}^{\rm con}$, $\Omega_{\beta,\lambda}^{\rm con}$, the von Neumann algebra $\mm_{\beta}^{\rm con}$ and the Liouville operators ${\cal L}_{0}$, ${\cal L}_{\lambda}$ are direct integrals over corresponding fibers (labelled by $\xi$ and weighted with the given measure $\mu$). \\
\indent
We are thus in the setting of Definition 2' with the following identifications: $\Omega_\xi$ is the fiber of $\Omega_{\beta,\lambda}^{\rm con}$, $\pi_\xi$ is the fiber of $\pi$ and  $\sigma_t^\xi$ is the $*$automorphism group $e^{itL_{\lambda,\xi}}(\cdot) e^{-itL_{\lambda,\xi}}$ of $\pi_\xi({\frak A})''$, where $L_{\lambda,\xi}$ is the fiber of ${\cal L}_\lambda$. Our {\it weak coupling} result on Return of Equilibrium reads: for all $\mu,A,B$
\begin{equation}
\lim_{\lambda\rightarrow 0}\lim_{T\rightarrow\infty}\frac 1T\int_0^T \!\!dt\,\omega^\mu(B^*\sigma_\lambda^t(A)B)=\int_{\r^2}d\mu(\xi)\, \omega_{\beta,0}^\xi(B^*B)\omega_{\beta,0}^\xi(A),
\label{intro146'}
\end{equation}
where $\omega_{\beta,0}^\xi$ is the state determined by $\Omega_{\beta,0}^\xi$, and $\sigma_\lambda^t$ is the interacting dynamics generated by ${\cal L}_\lambda$. The expression $\sigma_\lambda^t(A)$ has to be understood {\it cum grano salis}, in the sense of Definition 2' (see also \fer{cumgranosalis}). \\
\indent
The result \fer{intro146'} follows from our Theorem \ref{stabthm}: for $\epsilon>0$, $\xi\in[\rhocrit,\infty)\times S^1$, $B\in\frak A$ fixed, there is a $\lambda_0>0$ s.t. if $0<|\lambda|<\lambda_0$ then 
\begin{equation}
\left|
\lim_{T\rightarrow\infty} \frac 1T\!\int_0^T \!\! \omega_{\beta,\lambda}^\xi\big(B^*\sigma_{\lambda,\xi}^t(A)B\big)dt   
-\omega_{\beta,\lambda}^\xi(B^*B)\omega_{\beta,\lambda}^\xi(A)\right|
<\epsilon \|A\|,
\label{intro146}
\end{equation}
for all $A\in\frak A$, and where $\omega_{\beta,\lambda}^\xi$ is the vector state given by $\Omega_{\beta,\lambda}^\xi$. 
We expect that Return to Equilibrium should hold in the sense of Definition 3' but cannot prove it due to reasons we outline in Subsection \ref{subsubdiscussion}.\\
\indent
The proof of Theorem \ref{stabthm} is based on our Theorem \ref{chev} which describes the structure $\ker L_{\lambda,\xi}$, which in turn follows from a positive commutator argument combined with a new Virial Theorem, Theorem \ref{vthm}.\\

We prove \fer{intro146} under a condition of regularity and ``effectiveness'' of the interaction. Let us close this section by discussing the physical meaning of the latter condition. 
Consider first the Bose gas at critical density $\rhocrit(\beta)$ for some fixed temperature $1/\beta$ (so that there is no condensate, $\rho_0=0$). Heuristically, the probability of trapping a Boson in a state $f$ in the quantum dot is given by
\begin{equation}
\left| \scalprod{G_+\otimes a(f)\varphi\otimes\widetilde\Omega}{e^{-itH_\lambda}\varphi\otimes\widetilde\Omega}\right|^2,
\label{tr1}
\end{equation}
where $\varphi$ an eigenstate of the quantum dot Hamiltonian and the Bose gas is in the equilibrium state $\widetilde\Omega$ (for the calculation, we put the system in a box and $\widetilde\Omega$ is a vector in Fock space with Bosons distributed according to a discrete distribution approaching the Planck distribution as the box size increases). The interacting Hamiltonian is $H_\lambda=H_0+\lambda (G_+\otimes a(g) + G_-\otimes a^*(g))$. The second order contribution in $\lambda$ to \fer{tr1}, for large values of $t$, is 
\begin{equation}
  P_2 = C\frac{\lambda^2}{(e^{\beta\omega(1)} -1)^2}|f(1)g(1)|^2,
\label{tr2}
\end{equation}
where we assume that $f(r),g(r)$ are radially symmetric, and where $C$ is a constant independent of $\beta, f,g$. $P_2$ gives the probability of the second order process where a Boson gets trapped in the quantum dot; the excitation energy is $1$ (the quantum dot Hamiltonian \fer{r1} has equidistant eigenvalues) and the probability density of finding a Boson with energy $\omega(1)=1$ per unit volume is $\propto(e^\beta-1)^{-1}$, according to \fer{115}. In order not to suppress this trapping process at second order in the coupling constant we assume that $g(1)\neq 0$ (``effective coupling''). \\
\indent
Next let us investigate the influence of the condensate. For this we fix a density $\rho_0$ of the Bose gas and consider very low temperatures ($\beta\rightarrow\infty)$, so that most particles are in the condensate. If $\widetilde\Omega$ denotes the corresponding state of the Bose gas then we calculate the second order in $\lambda$ of \fer{tr1} to be
\begin{equation}
Q_2(t)=C(1-\cos t)\lambda^2\rho_0^2|f(0)g(0)|^2.
\label{tr3}
\end{equation}
We see from \fer{tr3} that if $g(0)=0$ then there is no coupling to the modes of the condensate: a physically trivial situation where the condensate evolves freely and the small system coupled to the ``excited modes'' undergoes return to equilibrium. In this paper we develop a theory which includes the case $g(0)\neq 0$, a situation which could not be handled by approaches developped so far.

\section{Definition of the model and main results}
\label{mainresultsection}

In Section \ref{subsdefmod} we introduce the class of systems considered in this paper and  we explain the central decomposition of the equilibrium state with a condensate (references we find useful for this are [AW], [LP], [C] and also [H]). Our main results are presented in Section \ref{mainresults}, at the end of which we also give the quite short proof of the stability result, Theorem \ref{stabthm}.

\subsection{Definition of model}
\label{subsdefmod}

We introduce the uncoupled system in Subsection \ref{subsub1} and present its Hilbert space (GNS) description (see \fer{141.1}, \fer{141'}) including the uncoupled standard Liouvillian ${\cal L}_0$,  \fer{136}. The interaction is defined by an interacting standard Liouvillian ${\cal L}_\lambda$, introduced in Subsection \ref{intsyst}, \fer{152}.

\subsubsection{Non-interacting system}
\label{subsub1}

The {\it states} of the small system are determined by density matrices $\rho$ acting on $\cx^d$.
A density matrix is a positive trace-class operator, normalized as $\tr\rho=1$, and the corresponding state
\begin{equation}
\omega_\rho(A)=\tr(\rho A),\ \ \ A\in{\cal B}(\cx^d)
\label{102}
\end{equation}
is a normalized positive linear functional on the $C^*$-algebra ${\cal B}(\cx^d)$ of all bounded operators on $\cx^d$, which we call the algebra of {\it observables}. The (Heisenberg-) {\it dynamics} of the small system is given by the group of $*$auto\-morphisms of ${\cal B}(\cx^d)$ generated by the {\it Hamiltonian} $H_1$ given in \fer{r1}, 
\begin{equation}
\alpha_1^t(A)=e^{itH_1} A e^{-itH_1},\ \ \ t\in\r.
\label{103}
\end{equation}
Denote the normalized eigenvector of $H_1$ corresponding to the eigenvalue $E_j$ by $\varphi_j$. 
Given any inverse temperature $0<\beta<\infty$ the {\it Gibbs state} $\omega_{1,\beta}$ is the unique $\beta$-KMS state on ${\cal B}(\cx^d)$ associated to the dynamics \fer{103}. It corresponds to the density matrix
\begin{equation}
\rho_\beta=\frac{e^{-\beta H_1}}{\tr e^{-\beta H_1}}.
\label{105}
\end{equation}
\indent
Let $\rho$ be any density matrix of rank $d$ (equivalently, $\rho>0$) and let $\{\varphi_j\}_{j=0}^{d-1}$ be an orthonormal basis of eigenvectors of $\rho$, corresponding to eigenvalues $0< p_j <1$, $\sum_j p_j=1$. The GNS representation of the pair $({\cal B}(\cx^d),\omega_\rho)$ is given by $(\h_1,\pi_1,\Omega_1)$, where the Hilbert space $\h_1$ and the cyclic (and separating) vector $\Omega_1$ are  
\begin{eqnarray}
\h_1&=&\cx^d\otimes\cx^d,
\label{105.2}\\
\Omega_1&=&\sum_j \sqrt{p_j}\  \varphi_j\otimes \varphi_j \in \cx^d\otimes\cx^d,
\label{105.1}
\end{eqnarray}
and the representation map $\pi_1:{\cal B}(\cx^d)\rightarrow {\cal B}(\h_1)$ is
\begin{equation}
\pi_1(A)= A\otimes \bbbone.
\label{105.2'}
\end{equation}
We introduce the von Neumann algebra
\begin{equation}
\mm_1={\cal B}(\cx^d)\otimes\bbbone_{\cx^d} \subset {\cal B}(\h_1).
\label{mm1}
\end{equation}
The modular conjugation operator $J_1$ associated to the pair $(\mm_1,\Omega_1)$ is given by 
\begin{equation}
J_1 \psi_{\ell}\otimes\psi_{r}= {\cal C}_1\psi_{r}\otimes {\cal C}_1\psi_{\ell},
\label{105.3}
\end{equation}
where ${\cal C}_1$ is the antilinear involution ${\cal C}_1\sum_j z_j\varphi_j =\sum_j \overline{z_j}\varphi_j$ (complex conjugate). According to \fer{105.1} and \fer{105} the vector $\Omega_{1,\beta}$ representing the Gibbs state $\omega_{1,\beta}$ is given by
\begin{equation}
\Omega_{1,\beta}
=\frac{1}{\sqrt{\tr e^{-\beta H_1}}} \sum_j e^{-\beta E_j/2}\varphi_j\otimes\varphi_j\in\h_1.
\label{gibbsvector}
\end{equation}

We now turn to the description of heat bath. Its algebra of observables is the {\it Weyl algebra} ${\frak W}(\dom)$ over some linear subspace of test functions $\dom\subset L^2(\r^3,d^3k)$. The elements of $\dom$ represent the wave functions of a single quantum particle of the heat bath. The choice of $\dom$ depends on the physics one wants to describe -- in particular, it is not the same for a system of Bosons with and without a condensate, as we will see shortly. For fixed $\dom$, ${\frak W}(\dom)$ is the $C^*$-algebra generated by elements $W(f)$, $f\in\dom$, called the Weyl operators, which satisfy the CCR \fer{106}.
 The $*$operation of ${\frak W}(\dom)$ is given by $W(f)^*=W(-f)$. The dynamics of the heat bath is described by the group of $*$automorphisms of ${\frak W}(\dom)$
\begin{equation}
\alpha^t_2(W(f))= W(e^{ith}f),
\label{108}
\end{equation}
where $h$ is a selfadjoint operator on $L^2(\r^3,d^3k)$. In the present paper, we choose $h$ to be the operator of multiplication by the function $\omega(k)$, see \fer{109}.  Our methods can be modified to accomodate for other dispersion relations than \fer{109}. \\
\indent
We choose the test function space $\dom$ to consist of all functions $f\in L^2(\r^3,d^3k)$ s.t. $E^{\rm con}_{\beta,\overline\rho}(f)$ exists (see \fer{r8}). 
If $\rho_0=0$ the r.h.s. of \fer{r8} reduces to the product of the first two exponentials (one may then extend $\dom$ to $L^2(\r^3, (1+\rho)d^3k)$), and if in addition $\beta\rightarrow\infty$ ($\Rightarrow\rho(k)=0$) then $E(f)=e^{-\frac{1}{4}\|f\|^2}$ is just the Fock generating functional corresponding to the zero temperature equilibrium state (in this case one may extend the test function space to all of $L^2(\r^3,d^3k)$). \\
\indent
The GNS representation of the pair $({\frak W}(\dom),\omega_{2,\beta})$, where $\omega_{2,\beta}$ is the equilibrium state with supercritical density $\overline\rho$ (determined by $E^{\rm con}_{\beta,\overline\rho}(f)$, \fer{r8}), has been given in [LP] as the triple $(\h_2, \pi_2,\Omega_2)$, where the representation Hilbert space is 
\begin{equation}
\h_2=\ff\otimes\ff\otimes L^2(\r^2,d\mu_{\beta,\overline\rho}),
\label{118}
\end{equation}
where $\ff=\ff(L^2(\r^3,d^3k))$ is the Bosonic Fock space over $L^2(\r^3,d^3k)$, and the probability measure $d\mu_{\beta,\overline\rho}$ is given in \fer{mubetarho}. The cyclic vector is 
\begin{equation}
\Omega_2=\Omegaf\otimes\Omegaf\otimes \1
\label{119}
\end{equation}
where $\Omegaf$ is the vacuum in $\ff$ and $\1$ is the normalized constant function in $L^2(\r^2,d\mu_{\beta,\overline\rho})$. The representation map $\pi_2:{\frak W}(\dom)\rightarrow {\cal B}(\h_2)$ is given by 
\begin{equation}
\pi_2(W(f))=\Wf(\sqrt{1+\rho}f)\otimes \Wf(\sqrt{\rho}\overline f)\otimes e^{-i\Phi(f,\xi)},
\label{120}
\end{equation}
where  
$
\Wf=e^{i\varphi_{\!\ff}(f)}
$ 
is a Weyl operator in Fock representation, the field operator $\varphif(f)$ is 
\begin{equation}
\varphif(f)=\frac{1}{\sqrt 2}(\af^*(f)+\af(f))
\label{fieldinfock}
\end{equation}
and $\af^*(f)$, $\af(f)$ are the smeared out creation, annihilation operators satisfying the commutation relations 
\begin{equation}
[\af(f),\af^*(g)]=\scalprod{f}{g},\  [\af(f),\af(g)]=[\af^*(f),\af^*(g)]=0. 
\label{atorsinfock}
\end{equation}
Our convention is that $f\mapsto \af(f)$ is an antilinear map. The phase $\Phi\in\r$ is given by
\begin{equation}
\Phi(f,\xi)= (2\pi)^{-3/2}\sqrt{2(r-\rhocrit)}\Big( (\RE f(0))\cos\theta +(\IM f(0))\sin\theta\Big),
\label{121}
\end{equation}
for $\xi=(r,\theta)\in [\rhocrit,\infty)\times S^1$. 
In the absence of a condensate the third factor in \fer{118}, \fer{119} and \fer{120} disappears and the representation reduces to the ``Araki-Woods representation'' in the form it has appeared in a variety of recent papers. We denote this representation by $\pi_{0}$. More precisely, let $\pi_0:{\frak W}(\dom)\rightarrow{\cal B}(\ff\otimes\ff)$ be the representation
\begin{equation}
\pi_0(W(f))= \Wf(\sqrt{1+\rho}f)\otimes \Wf(\sqrt\rho\overline f).
\label{132'}
\end{equation}
It is well known that the vector
\begin{equation}
\Omega_0= \Omegaf\otimes \Omegaf
\label{133'}
\end{equation}
is cyclic and separating for the von Neumann algebra 
\begin{equation}
\mm_0:=\pi_0({\frak W}(\dom))'' \  \subset {\cal B}(\ff\otimes\ff)
\label{130'}
\end{equation}
(it defines a KMS state w.r.t. the free field dynamics). We also introduce the von Neumann algebra 
\begin{equation}
\mm_2:= \pi_2({\frak W}(\dom))'' =\mm_0\otimes{\cal M} \subset {\cal B}(\h_2),
\label{130}
\end{equation}
where $\cal M$ is the abelian von Neumann algebra of all multiplication operators on $L^2(\r^2,d\mu_{\beta,\overline\rho})$. It satisfies ${\cal M}'=\cal M$. The  equality in \fer{130} follows from this: clearly we have ${\mm_0}' \otimes{\cal M}\subset {\mm_2}'$, so taking the commutant gives 
$
\mm_0\otimes{\cal M}\supset \mm_2
$. 
The reverse inclusion is obtained from $\bbbone_{\ff\otimes\ff}\otimes {\cal M}\subset\mm_2$ and $\mm_0\otimes\bbbone_{L^2(\r^2)}\subset \mm_2$ (see [LP]). \\
\indent
It is well known that the von Neumann algebra $\mm_0$, \fer{130'}, is a factor. That means  its center is trivial, ${\frak Z}(\mm_0)=\mm_0\cap{\mm_0}'\cong\cx$. However, we have ${\frak Z}(\mm_2)=(\mm_0\otimes{\cal M})\cap ({\mm_0}'\otimes {\cal M})$, i.e.
\begin{equation}
{\frak Z}(\mm_2)= \bbbone_{\ff\otimes\ff}\otimes{\cal M},
\label{132}
\end{equation}
so the von Neumann algebra $\mm_2$ is {\it not a factor}. This is clear on physical grounds since the system has long-range correlations, see Section \ref{fbg}. One can decompose $\mm_2$ into a direct integral of factors. The Hilbert space \fer{118} is the direct integral
\begin{equation}
\h_2=\dirint_{\r^2}d\mu_{\beta,\overline\rho}(\xi) \ \ff\otimes\ff,
\label{135'}
\end{equation}
and the formula (see \fer{119}, \fer{120}, \fer{132'},  \fer{133'}) 
\begin{equation}
\omega_{2,\beta}(W(f))=\scalprod{\Omega_2}{\pi_2(W(f))\Omega_2}=\int_{\r^2}
d\mu_{\beta,\overline\rho}(\xi) e^{-i\Phi(f,\xi)} \scalprod{\Omega_0}{\pi_0(W(f))\Omega_0}
\label{133}
\end{equation}
shows that $\pi_2$ is decomposed as 
\begin{equation}
\pi_2=\dirint_{\r^2}d\mu_{\beta,\overline\rho}(\xi)\ \pi_\xi,
\label{136.1}
\end{equation}
where $\pi_\xi: {\frak W}(\dom)\rightarrow {\cal B}(\ff\otimes\ff)$ is the representation defined by 
\begin{equation}
\pi_\xi(W(f)) = e^{-i\Phi(f,\xi)}\pi_0(W(f)).
\label{134}
\end{equation}
For each  fixed $\xi$,
\begin{equation}
\pi_\xi({\frak W}(\dom))'' =\mm_0
\label{''}
\end{equation}
is a factor. Accordingly we have
\begin{equation}
\mm_2=\dirint_{\r^2}d\mu_{\beta,\overline\rho}(\xi)\ \mm_0.
\label{135}
\end{equation}
The GNS representation of $({\frak A},\ocond_{\beta,0})$ (see \fer{intro122}) is just $(\h,\pi,\Omega)$, where
\begin{eqnarray}
\h&=& \h_1 \otimes \h_2\nonumber \\
\pi&=& \pi_1\otimes\pi_2\label{141.1}\\
\Obocond&=& \Omega_{1,\beta}\otimes\Omega_2.\label{141'}
\end{eqnarray}
The free dynamics is given by the group of $*$automorphisms $\alpha^t_0$, \fer{124}. Let 
\begin{equation}
\mbcond:=\pi({\frak A})''=\mm_1\otimes\mm_2 =\dirint_{\r^2} d\mu_{\beta,\overline\rho}(\xi) \ \mm_1\otimes\mm_0 \ \ \subset {\cal B}(\h)
\label{125}
\end{equation}
be the von Neumann algebra obtained by taking the weak closure of all observables of the combined system, when represented on $\h$. 
To see how we can implement the uncoupled dynamics in $\h$ we use that (for all $t\in\r$) $\Phi(e^{i\omega t}f,\xi)=\Phi(f,\xi)$, which follows from $\omega(0)=0$, see \fer{121} and \fer{109}. Thus  
\begin{equation}
\pi_2(\alpha_2^t(W(f)))=\dirint_{\r^2} d\mu_{\beta,\overline\rho}(\xi) 
e^{-i\Phi(f,\xi)} \pi_0(W(e^{i\omega t}f)).
\label{137a}
\end{equation}
It is well known and easy to verify that for $A\in{\frak A}$,
\begin{equation}
(\pi_1\otimes \pi_0)(\alpha_0^t(A))= e^{itL_0}(\pi_1\otimes\pi_0)(A)e^{-itL_0},
\label{139a}
\end{equation}
where the selfadjoint operator $L_0$ on $\h_1\otimes\ff\otimes\ff$ is given by 
\begin{eqnarray}
L_0&=&L_1+L_2,\label{lnot}\\
L_1&=&H_1\otimes \bbbone_{\cx^d} -\bbbone_{\cx^d}\otimes H_1, \label{128} \\
L_2 &=&\d\Gamma(\omega)\otimes\bbbone_\ff -\bbbone_\ff\otimes\d\Gamma(\omega). \label{129}
\end{eqnarray}
Here $\d\Gamma(\omega)$ is the second quantization of the operator of multiplication by $\omega$ on $L^2(\r^3,d^3k)$. 
We will omit trivial factors $\bbbone$ or indices $_{\cx^d}$, $_\ff$ whenever we have the reasonable hope that no confusion can arise (e.g. $L_1$ really means $L_1\otimes\bbbone_\ff\otimes\bbbone_\ff$). It follows from \fer{125}--\fer{129} that the uncoupled dynamics $\alpha_0^t$ is unitarily implemented in $\h$ by 
\begin{equation}
\pi(\alpha_0^t(A)) =e^{it\L_0}\pi(A)e^{-it\L_0},
\label{126}
\end{equation}
where the {\it standard, non-interacting Liouvillian} $\L_0$ is the selfadjoint operator on $\h$ with constant ($\xi$-independent) fiber $L_0$,
\begin{equation}
\L_0=\dirint_{\r^2} d\mu_{\beta,\overline\rho}(\xi)\ L_0.
\label{136}
\end{equation}
The r.h.s. of \fer{126} extends to a $*$automorphism group $\sigma^t_0$ of $\mbcond$ which is reduced by the decomposition \fer{125}. We write
\begin{equation}
\sigma_0^t = \dirint_{\r^2} d\mu_{\beta,\overline\rho}(\xi)\ \sigma_{0,\xi}^t,
\label{137}
\end{equation}
where $\sigma_{0,\xi}^t$ is the $*$automorphism group of $\mm_1\otimes\mm_0$ generated by $L_0$. As is well known,
\begin{equation}
\Omega_{\beta,0}=\Omega_{1,\beta}\otimes\Omega_0
\label{kmsbo}
\end{equation}
is a $(\beta,\sigma_{0,\xi}^t)$-KMS state of $\mm_1\otimes\mm_0$. The modular conjugation operator $J$ associated to $(\mm_0,\Omega_{1,\beta}\otimes\Omega_0)$ is 
\begin{equation}
J=J_1\otimes J_0,
\label{138}
\end{equation}
where $J_1$ is given by \fer{105.3} and where the action of $J_0$ on $\ff\otimes\ff$ is determined by antilinearly extending the relation
\begin{equation}
J_0\pi_0(W(f))\Omega_0= \Wf(\sqrt\rho f)\otimes\Wf(\sqrt{1+\rho}\, \overline{f})\Omega_0.
\label{139}
\end{equation}
$J_0$ defines an antilinear representation of the Weyl algebra according to $W(f)\mapsto J_0 \pi_0(W(f)) J_0$, which commutes with the representation $\pi_0$ given in \fer{132'}. We view this as a consequence of the Tomita-Takesaki theory which asserts that ${\mm_0}'=J_0\mm_0 J_0$. \\
\indent
It follows from \fer{141'}, \fer{125}, \fer{137} that 
\begin{equation}
\Obocond =\dirint_{\r^2} d\mu_{\beta,\overline\rho}(\xi)\ \Omega_{\beta,0}
\label{140}
\end{equation}
is a $(\beta,\sigma_0^t)$-KMS state on $\mbcond$, and that the modular conjugation operator $\cal J$ associated to $(\mbcond,\Obocond)$ is given by
\begin{equation}
{\cal J}=\dirint_{\r^2} d\mu_{\beta,\overline\rho}(\xi) \ J_1\otimes J_0.
\label{141}
\end{equation}
The standard Liouvillian $\L_0$, \fer{136}, satisfies the relation
\begin{equation}
{\cal J}\L_0 =-\L_0{\cal J}.
\label{142}
\end{equation}
One can choose different generators to implement the dynamics $\alpha_0^t$ on $\h$ (by adding to the standard $\L_0$ any selfadjoint element affiliated with the commutant $(\mbcond)'$\,). The choice \fer{136} is compatible with the symmetry $\mbcond\cong (\mbcond)'$, in that it also implements $\alpha_0^t$ for the antilinear representation ${\cal J}\pi(\cdot){\cal J}$. Another way to say this is that the standard Liouvillian \fer{136} is the only generator which implements the non-interacting dynamics $\alpha_0^t$ and satisfies
\begin{equation}
\L_0\Obocond=0,
\label{143}
\end{equation}
see e.g. [BR, DJP].

\subsubsection{Interacting system}
\label{intsyst}

We define the coupled dynamics, i.e. the interaction between the small system and the Bose gas, by specifying a $*$automorphism group $\sigma^t_{\lambda}$ of the von Neumann algebra $\mbcond$ (the ``perturbed'' or ``interacting dynamics''). One may argue that a conceptually more satisfying way is to introduce a representation independent {\it regularized} dynamics as a $*$automorphism group of $\frak A$ and then to remove the regularization once the dynamics is represented on a Hilbert space. This procedure can be implemented by following the arguments of [FM1], where it has been carried out for the Bose gas without condensate. The resulting dynamics is of course the same for both approaches. For a technically more detailed exposition of the following construction we refer the reader to [FM1].\\
\indent
The interaction between the two subsystems is given formally by \fer{144}, which we understand as an operator in a regular representation of the Weyl algebra, so that the creation and annihilation operators are well defined. We could treat interactions which are sums over finitely many terms of the form \fer{144}, simply at the expense of more complicated notation.\\
\indent
The field operator $\varphi(f)=\frac{1}{i}\partial_t|_{t=0}\pi(W(tf))$ in the representation $\pi$, \fer{141.1},  is easily calculated to be 
\begin{eqnarray}
\varphi(f)&=&\dirint_{\r^2} d\mu_{\beta,\overline\rho}(\xi)\  \varphi_\xi(f),\label{145}\\
\varphi_\xi(f) &=& \varphif(\sqrt{1+\rho}f)\otimes\bbbone +\bbbone\otimes\varphif(\sqrt\rho\overline{f}) -\Phi(f,\xi),
\label{146}
\end{eqnarray}
where $\Phi(f,\xi)$ is given in \fer{121}, and where $\varphif(f)$ is given in \fer{fieldinfock}. Define the interaction operator by 
\begin{eqnarray}
\lefteqn{
 V= G_+\otimes\bbbone_{\cx^d}\otimes \left( a_{\ff}\big(\sqrt{1+\rho}g\big)\otimes\bbbone_\ff +\bbbone_\ff\otimes a_\ff^*\big(\sqrt{\rho}\overline g\big)\right.} \nonumber\\
&&\left.-(2\pi)^{-3/2}\sqrt{2(r-\rhocrit)}\ \overline{g(0)} \ e^{i\theta}\right) + \rm{adjoint},
\label{147}
\end{eqnarray}
which corresponds formally to $\pi\big(G_+\otimes a(g)\big)+{\rm adjoint}$ (apply \fer{141.1} to \fer{144}). 
$V$ is an unbounded selfadjoint operator on $\h$ which is affiliated with $\mbcond$. For $t\in\r$, $A\in\mbcond$ we set
\begin{eqnarray}
\sigma_{\lambda}^t(A)&=&\sum_{n\geq 0} (i\lambda)^n\int_0^tdt_1\ldots\int_0^{t_{n-1}}\!\!\!dt_n \big[e^{it_n\L_0}Ve^{-it_n\L_0},\big[\cdots\nonumber\\
&&\ \ \ \ \ \cdots\big[ e^{it_1\L_0}Ve^{-it_1\L_0}, A\big]\cdots\big]\big].
\label{148}
\end{eqnarray}
The series converges in the strong sense on a dense set of vectors, for any $A\in\mbcond$, $\lambda,t\in\r$ (see e.g. [FM1]). Since $V$ is affiliated with $\mbcond$ and $e^{it\L_0}\cdot e^{-it\L_0}$ leaves $\mbcond$ invariant, the integrand in \fer{148} does not change if we add to each $e^{it_j\L_0}Ve^{-it_j\L_0}$ a term $-{\cal J} e^{it_j\L_0}Ve^{-it_j\L_0}{\cal J}=-e^{it_j\L_0}{\cal J} V{\cal J} e^{-it_j\L_0}$ (because this term is affiliated with the commutant $(\mbcond)'$). In other words, $V$ in \fer{148} can be replaced by $V-{\cal J}V{\cal J}$. The r.h.s. of \fer{148} is then identified as the Dyson series expansion of 
\begin{equation}
e^{it\L_\lambda} A e^{-it\L_\lambda},
\label{149}
\end{equation}
where the standard, interacting  Liouvillian $\L_\lambda$ is the selfadjoint operator 
\begin{equation}
\L_\lambda =\L_0 +\lambda (V-{\cal J}V{\cal J})\equiv \L_0 +\lambda{\cal I}.
\label{150}
\end{equation}
Subtracting the term ${\cal J}V{\cal J}$ serves to preserve the symmetry \fer{142} when adding the perturbation, i.e., we have ${\cal J}\L_\lambda=-\L_\lambda{\cal J}$. It is not hard to verify that \fer{149} defines a $*$automorphism group
\begin{equation}
\sigma^t_\lambda(A)=e^{it\L_\lambda} A e^{-it\L_\lambda}
\label{151}
\end{equation}
of $\mbcond$ which defines the interacting dynamics. 
The Liouvillian $\L_\lambda$ is reduced by the direct integral decomposition,
\begin{eqnarray}
\L_\lambda &=&\dirint_{\r^2} d\mu_{\beta,\overline\rho}(\xi) \ \Llt,
\label{152}\\
\Llt&=& L_0 +\lambda I_\xi,
\label{151.2}
\end{eqnarray}
where $L_0$ is given in \fer{lnot} and where we define
\begin{eqnarray}
I_\xi &=& I +K_\xi,\label{153}\\
I&=& G_+\otimes\bbbone_{\cx^d}\otimes\left\{ \af(\sqrt{1+\rho}\, g)\otimes\bbbone_\ff +\bbbone_\ff\otimes \af^*(\sqrt\rho\, \overline{g})\right\}+\mbox{adj.}\nonumber\\
&& \hspace*{-1cm}- \bbbone_{\cx^d}\otimes \cc_1G_+\cc_1\otimes\left\{ \af^*(\sqrt\rho g)\otimes\bbbone_\ff +\bbbone_\ff\otimes \af(\sqrt{1+\rho}\, \overline{g})\right\}+\mbox{adj.}
\label{153.1}\\
K_\xi&=& K_\xi^1\otimes\bbbone_{\cx^d}\otimes\bbbone_{\ff\otimes\ff} -\bbbone_{\cx^d}\otimes \cc_1 K_\xi^1\cc_1\otimes \bbbone_{\ff\otimes\ff}
\label{here}\\
K^1_\xi&=& -2(2\pi)^{-3/2} \sqrt{r-\rhocrit}\left( G_+ \overline{g(0)}e^{i\theta} +G_- g(0)e^{-i\theta}\right)
\label{153.153}
\end{eqnarray}
with $\cc_1$ given in \fer{105.3}, and where the creation and annihilation operators $\af^*$, $\af$ are defined by \fer{atorsinfock}. 
It is convenient to write (compare with \fer{137})
\begin{equation}
\sigma_\lambda^t =\dirint_{\r^2} d\mu_{\beta,\overline\rho}(\xi) \ \sigma_{\lambda,\xi}^t,
\label{151.1}
\end{equation}
where $\sigma_{\lambda,\xi}^t$ is the $*$automorphism group on $\mm_1\otimes \mm_0$ generated by $\Llt$, \fer{151.2}.\\
\indent
To the interacting dynamics \fer{151} corresponds a $\beta$-KMS state on $\mbcond$, the equilibrium state of the interacting system. It is given by the vector
\begin{equation}
\Oblcond=\dirint_{\r^2} d\mu_{\beta,\overline\rho}(\xi) \ \Omega_{\beta,\lambda}^\xi,
\label{kms1}
\end{equation}
where
\begin{equation}
\Omega_{\beta,\lambda}^\xi=(Z_{\beta,\lambda}^\xi)^{-1} e^{-\beta(L_0+\lambda I_{\xi,\ell})/2}\Omega_{\beta,0}
\in\h_1\otimes\ff\otimes\ff,
\label{154}
\end{equation}
with a normalization factor $Z_{\beta,\lambda}^\xi$ ensuring that $\|\Omega_{\beta,\lambda}^\xi\|=1$, and where 
$I_{\xi,\ell}$ is obtained by dropping the second term (the one coming with a minus sign) both in \fer{153.1} and in \fer{here}. The fact that $\Omega_{\beta,0}$, \fer{kmsbo}, is in the domain of the unbounded operator $e^{-\beta(L_0+\lambda I_{\xi,\ell})/2}$, provided 
\begin{equation}
\|g/\sqrt\omega\|_{L^2(\r^3)}<\infty,
\label{n2}
\end{equation}
can be seen by expanding the exponential in a Dyson series and verifying that the series applied to $\Omega_{\beta,0}$ converges, see e.g. [BFS]. It then follows from the generalization of Araki's perturbation theory of KMS states, given in [DJP], that $\Omega_{\beta,\lambda}^\xi$ is a $(\beta,\sigma^t_{\lambda,\xi})$-KMS state on $\mm_1\otimes \mm_0$, and that
\begin{equation}
\Llt \, \Omega_{\beta,\lambda}^\xi =0.
\label{155}
\end{equation}
We conclude that $\Oblcond$ is a $(\beta,\sigma_\lambda^t)$-KMS state on $\mbcond$, and that $\L_\lambda \Oblcond=0$.

\subsection{Main results}
\label{mainresults}

We make two assumptions on the  form factor $g$ determining the interaction (see \fer{144}, \fer{147}).

\begin{enumerate}
\item[(A1)]
{\it Regularity.\ } The form factor $g$ is a function in $C^4(\r^3)$ and satisfies 
\begin{equation*}
\| (1+1/\sqrt{\omega})(k\cdot\nabla_k)^j g\|_{L^2(\r^3,d^3k)}<\infty,
\end{equation*}
for $j=0,\ldots,4$, and $\|\, (1+\omega)^2 g\|_{L^2(\r^3,d^3k)}<\infty$. 

\item[(A2)]
{\it Effective coupling.\ } We assume that $\int_{S^2}d\sigma\ |g(1,\sigma)|^2\neq 0$. Here, $g$ is represented in polar coordinates. 
\end{enumerate}
{\it Remarks.\ } 1)\ Condition (A1) is used in the application of the virial theorem -- we choose the generator of dilations $\frac{1}{2}(k\cdot\nabla_k+\nabla_k\cdot k)$ to be the conjugate operator in the theory.\\
\indent
2)\ Condition (A2) is often called the {\it Fermi Golden Rule Condition}. It guarantees that the processes of absorption and emission of field quanta by the small system, which are the origin of the stability of the equilibrium, are effective, see the discussion in the introduction. In a situation where the spectrum of $H_1$ has gaps $\Delta_j$ between neighbouring eigenvalues, condition (A2) is replaced by $\min_j\int_{S^2} d\sigma |g(\omega^{-1}(\Delta_j),\sigma)|^2\neq 0$. This means that the form factor should couple the modes of the gas which induce transitions of the small system.

\begin{theorem}{\bf (Weak coupling return to equilibrium).\ }
\label{stabthm}
Assume conditions (A1) and (A2). Let $\epsilon>0$, $\xi\in [\rhocrit,\infty)\times S^1$, $B\in\frak A$ be fixed. There is a $\lambda_0(\epsilon,\xi,B)>0$ s.t. if $0<|\lambda|<\lambda_0(\epsilon,\xi,B)$ then 
\begin{multline}
\left|
\lim_{T\rightarrow\infty}\frac 1 T\int_0^T \! dt\,
\scalprod{\pi_\xi(B)\Omega_{\beta,\lambda}^\xi}{\sigma^\xi_{t,\lambda}(\pi_\xi(A))\pi_\xi(B)\Omega_{\beta,\lambda}^\xi}
-\omega_{\beta,\lambda}^\xi(B^*B)\omega_{\beta,\lambda}^\xi(A)\right|\\
<\epsilon \|A\|,
\label{158}
\end{multline}
for all $A\in\frak A$. The coupled KMS state $\omega_{\beta,\lambda}^\xi$ is determined by the vector \fer{154}. Set  
\begin{equation}
\omega^\mu(B^*\sigma_\lambda^t(A)B) := \int_{\r^2}d\mu(\xi) \scalprod{\pi_\xi(B)\Omega_{\beta,\lambda}^\xi}{\sigma_{t,\lambda}^\xi(\pi_\xi(A)) \pi_\xi(B)\Omega_{\beta,\lambda}^\xi},
\label{cumgranosalis}
\end{equation}
where $\mu$ is any probability measure supported on $[\rhocrit,\infty)\times S^1$. If follows from \fer{158} that 
\begin{equation}
\lim_{\lambda\rightarrow 0}\lim_{T\rightarrow 0}\frac 1T\int_0^T\!\!dt\,   \omega^\mu(B^*\sigma_\lambda^t(A)B) =\int_{\r^2}d\mu(\xi)\, \omega_{\beta,0}(B^*B)\omega_{\beta,0}(A),
\end{equation}
where $\omega_{\beta,0}$ is the uncoupled KMS state determined by the vector $\Omega_{\beta,0}$, \fer{kmsbo}. 
\end{theorem}

{\it Remark.\ } We expect the r.h.s. of \fer{158} to be zero for $\lambda$ sufficiently small. This has been proven to hold for systems without a condensate (with varying conditions on the interaction and varying modes of convergence) in several papers, see [JP1, BFS, M1, DJ, FM2]. The obstruction to applying the strategies of these papers is that they all need the condition that either $g(0)=0$, or $g(k)\sim |k|^{-1/2}$, as $|k|\rightarrow 0$. The first case is uninteresting in the presence of a condensate (no coupling to the modes of the condensate!), and the second type of form factor does not enter into the description of a system with a condensate (see \fer{121}). We refer to Subsection \ref{subsubdiscussion} for a more detailed discussion of this point.\\
\indent
In order to state the virial theorem and to measure regularity of eigenvectors of $\Llt$, \fer{151.2}, we introduce the non-negative selfadjoint operator 
\begin{equation}
\Lambda =\d\Gamma(\omega)\otimes\bbbone_\ff +\bbbone_\ff\otimes\d\Gamma(\omega), 
\label{202}
\end{equation}
where $\d\Gamma(\omega)$ is the second quantization of the operator of multiplication by $\omega(k)$ on $L^2(\r^3, d^3k)$, c.f. \fer{109}. 
The kernel of $\Lambda$ is spanned by the vector $\Omega_0=\Omega_\ff\otimes\Omega_\ff$ (c.f. \fer{133'}) and $\Lambda$ has no nonzero eigenvalues. The operator $\Lambda$ represents the quadratic form $i[L_0,A]$, the commutator of $L_0$ with the {\it conjugate operator}
\begin{equation}
A=\d\Gamma(a_{\rm d})\otimes\bbbone_\ff -\bbbone_\ff\otimes \d\Gamma(a_{\rm d}),
\label{202.2}
\end{equation}
where $a_{\rm d}$ is the selfadjoint generator of dilations on $L^2(\r^3,d^3k)$,
\begin{equation}
a_{\rm d} = i\left( k\cdot\nabla_k +\frac 32\right).
\label{202.3}
\end{equation}
The formal relation $\Lambda=i[L_0,A]$ follows from $i[\omega,a_{\rm d}]=\omega$ (for relativistic Bosons, see \fer{109}; in the non-relativistic case we have $i[\omega,a_{\rm d}/2]=\omega$). 
The selfadjoint operator representing the quadratic form $i[\Llt,A]$ is easily calculated to be (see \fer{153.1}) 
\begin{eqnarray}
C_1&=&\Lambda +\lambda I_1\label{202.1}\\
I_1 &=& G_+\otimes\bbbone_{\cx^d}\otimes\left\{ \af(a_{\rm d}\sqrt{1+\rho}\, g)\otimes\bbbone_\ff -\bbbone_\ff\otimes \af^*(a_{\rm d}\sqrt\rho\, \overline{g})\right\}+\mbox{adj.}\nonumber\\
&& \hspace*{-1cm}- \bbbone_{\cx^d}\otimes \cc_1G_+\cc_1\otimes\left\{ \af^*(a_{\rm d}\sqrt\rho g)\otimes\bbbone_\ff -\bbbone_\ff\otimes \af(a_{\rm d}\sqrt{1+\rho}\, \overline{g})\right\}+\mbox{adj.}\nonumber\\
\label{202.4}
\end{eqnarray}
Similar expressions are obtained for the higher commutators of $\Llt$ with $A$, see Section \ref{vtsection}. Assumption (A1) guarantees that $(1+1/\sqrt{\omega})(a_{\rm d})^j\sqrt\rho\, g$ and $(1+1/\sqrt{\omega})(a_{\rm d})^j\sqrt{1+\rho}\, g$ are in $L^2(\r^3,d^3k)$, for $j=0,\ldots,4$, so the commutators of $\Llt$ with $A$, up to order four, are represented by selfadjoint operators (satisfying the technical requirements needed in the proof of the virial theorem).

\begin{theorem}{\bf (Virial Theorem, regularity of eigenvectors of $\Llt$).\ }
\label{vthm}
Assume condition (A1) and let $\xi\in [\rhocrit,\infty)\times S^1$ be fixed. If $\psi$ is an eigenfunction of $\Llt$ then $\psi$ is in the form domain of $C_1$, \fer{202.1}, and
\begin{equation}
\scalprod{\psi}{C_1\psi}=0.
\label{202.5}
\end{equation}
There is a constant $c$ which does not depend on $\psi, \xi$ nor on $\beta\geq\beta_0$, for any $\beta_0>0$ fixed, such that 
\begin{equation}
\|\Lambda^{1/2}\psi\|\leq c |\lambda|\, \|\psi\|.
\label{202.6}
\end{equation}
\end{theorem}

{\it Remarks.\ } 1)\ Relation \fer{202.5} seems ``obvious'' from a formal point of view, writing $C_1=i[\Llt,A]=i[\Llt-e,A]$, and using that $(\Llt-e)^*=\Llt-e$, where $\Llt\psi=e\psi$. A {\it proof} of \fer{202.5} is certainly not trivial, though, and considerable effort has been spent by many authors to establish ``Virial Theorems'' (see e.g. [ABG] and [GG] for an overview, and also [M1], [FM1] for approaches similar to ours). \\
\indent
2) The regularity bound \fer{202.6} follows easily from \fer{202.5} and \fer{202.1} and from the standard fact that $I_1$ is infinitesimally small relative to $\Lambda^{1/2}$ (Kato), so that $0=\scalprod{\psi}{C_1\psi}\geq (1-\epsilon)\scalprod{\psi}{\Lambda\psi}-\frac{\lambda^2}{\epsilon}c \|\psi\|^2$, for any $\epsilon>0$, for some constant $c$ independent of $\xi$ and $\beta$, as mentioned in the theorem. We refer for a more complete exposition of this to [FM1]. \\
\indent
We prove Theorem \ref{vthm} in Section \ref{tca} by showing that the hypotheses leading to Theorem \ref{virialthm}, a more general result, are satisfied in the present situation.  Our next result describes the structure of ${\rm ker}\Llt$. Let $P(\Lambda\leq x)$ stand for the spectral projection of $\Lambda$ onto the interval $[0,x]$. 

\begin{theorem}{\bf (Structure of the kernel of $\Llt$).\ }
\label{chev}
Assume Conditions (A1), (A2) and let $\xi\in [\rhocrit,\infty)\times S^1$ be fixed. There is a number $\lambda_0>0$ s.t. if $0<|\lambda|<\lambda_0$ then 
 any normalized $\psi \in \ker(\Llt)$ satisfies 
\begin{equation}
\|P_{1,\beta}P(\Lambda\leq|\lambda|)\psi\| \geq 1-O(\lambda^0),
\label{seiv}
\end{equation}
where $P_{1,\beta}$ is the projection onto $\cx\Omega_{1,\beta}$ (see \fer{gibbsvector}) and $O(\lambda^0)$ is a vector whose norm, which is independent of $\psi$, tends to zero in the limit $\lambda\rightarrow 0$ (uniformly in $\xi$ in any compact set and in $\beta\geq\beta_0$, for any $\beta_0>0$ fixed). The constant $\lambda_0$ is uniform in $\xi$ in any compact set, and in $\beta\geq \beta_0$, for any fixed $\beta_0>0$.
\end{theorem}

Our proof of this theorem, given in Section \ref{ptchev}, relies on a positive commutator estimate and Theorem \ref{vthm}.
Expansion \fer{seiv} implies that the only vector in the kernel of $\Llt$ which does not converge weakly to zero, as $\lambda\rightarrow 0$, is the interacting KMS state $\Omega_{\beta,\lambda}^\xi$, \fer{154}. This information on the kernel of $\Llt$ {\it alone} enters our proof of Theorem \ref{stabthm}.\\

\noindent
{\bf Corollary 2.4\ }
{\it Assume Conditions (A1) and (A2) and let $P^\xi_{\beta,\lambda}$ the projection onto the subspace spanned by the interacting KMS state $\Omega_{\beta,\lambda}^\xi$, \fer{154}. Let $\xi\in[\rhocrit,\infty)\times S^1$ be fixed. Any normalized element $\psi\in\ker(\Llt)\cap \big(\ran P^\xi_{\beta,\lambda}\big)^\perp$ converges weakly to zero, as $\lambda\rightarrow 0$.  The convergence is uniform in $\xi$ in any compact set and in $\beta\geq\beta_0$, for any $\beta_0>0$ fixed.}\\

We prove the corollary in Section \ref{ptchev}. The virial theorem we present in Section \ref{vtsection}, Theorem \ref{virialthm}, is applicable to systems without a condensate, in which case one is interested in form factors $g$ which have a singularity at the origin. Theorem \ref{virialthm} can handle a wide range of such singularities (see the remark after Theorem \ref{vthmnocond}) and is therefore relevant in the study of {\it return to equilibrium} and {\it thermal ionization} for systems without condensate, as will be explained in [FM3].

\begin{theorem}{\bf (Improved Virial Theorem for systems without condensate).\ }
\label{vthmnocond}
Let $L_\lambda$ be the Liouvillian of a system without condensate, $L_\lambda=L_0+\lambda I$ (i.e., $K_\xi=0$), see \fer{153} and suppose that the form factor $g$ is in $C^4(\r^3\backslash\{0\})$ and satisfies the condition
\begin{eqnarray}
(1+1/\sqrt{\omega})(a_{\rm d})^j \sqrt{1+\rho}\,g, \    (1+1/\sqrt{\omega})(a_{\rm d})^j \sqrt{\rho}\,g\  \in L^2(\r^3,d^3k),\label{n01}\\
(1+\omega)^2(a_{\rm d})^j \sqrt{1+\rho}\,g, \
 (1+\omega)^2(a_{\rm d})^j \sqrt{\rho}\,g\  \in L^2(\r^3,d^3k),
\label{n1}
\end{eqnarray}
for $j=0,\ldots,4$. Then the conclusions \fer{202.5}, \fer{202.6} of Theorem \ref{vthm} hold.
\end{theorem}

{\it Remark.\ } An admissible infrared behaviour of $g$ satisfying \fer{n01}, \fer{n1} is $g(k)\sim |k|^p$, as $|k|\sim 0$, with $p>-1/2$ for relativistic Bosons (c.f. \fer{109}). The range of treatable values of $p$ obtained in previous works, [M1,DJ,FM1,FM2], is $p=-1/2,1/2,3/2$, $p>2$. Theorem \ref{vthmnocond} fills in the gaps between the discrete values of these admissible $p$.\\
\indent
The proof of Theorem \ref{vthmnocond} is the same as the one of Theorem \ref{vthm}, see Section \ref{tca}. 

\begin{theorem}
\label{t'}
Assume the setting of Theorem \ref{vthmnocond}, that (A2) holds and that $|g(k)|\leq c|k|^p$, for $|k|<c'$, for some constants $c, c'$, and where $p>-1/2$ (for relativistic Bosons, and $p>0$ for nonrelativistic ones). There is a number $\lambda_0>0$ s.t. if $0<|\lambda|<\lambda_0$ then any normalized $\psi\in \ker(L_\lambda)$ satisfies
\begin{equation}
\| P_{1,\beta}P(\Lambda\leq |\lambda|)\psi\| \geq 1- O(\lambda^0),
\label{seiv'}
\end{equation}
where $O(\lambda^0)$ is a vector whose norm is independent of $\psi$ and vanishes in the limit $\lambda\rightarrow 0$ (uniformly in $\beta\geq\beta_0$, for any $\beta_0>0$ fixed). The constant $\lambda_0$ does not depend on $\beta\geq \beta_0$, for any fixed $\beta_0>0$.
\end{theorem}

We give the proof Theorem \ref{t'} together with the proof of Theorem \ref{chev} in Section \ref{ptchev}. 

\subsubsection{``Weak coupling stability'' v.s. ``return to equilibrium'', and relation with infrared regularity of the coupling}
\label{subsubdiscussion}

A central tool in our analysis of the time-asymptotic behaviour of the system is the virial theorem, whose use imposes regularity conditions on the interaction. In particular, we must be able to control the commutators of $\Llt$ with the conjugate operator $A$ of degree up to four (see Section \ref{substavt}). Depending on the choice of $A$ this will impose more or less restrictive requirements on the interaction. A convenient choice for $A$ is obtained by representing $\ff\otimes\ff \cong \ff(L^2(\r\times S^2, du\times d\sigma))$ and choosing $A=i\d\Gamma(\partial_u)$ ({\it translation generator}). This choice, introduced in [JP1], has proven to be very useful and was adopted in [M1, DJ, FM1, FMS, FM2]. The commutator of the non-interacting Liouvillian $L_0=\d\Gamma(u)$ with $A$ (multiplied by $i$) is just $N=\d\Gamma(\bbbone)$, the number operator in $\ff(L^2(\r\times S^2, du\times d\sigma))$, which has a one-dimensional kernel and a {\it spectral gap} at zero. We may explain the usefulness of the gap as follows. If one carries out the proof of Theorem \ref{chev} with the translation generator as the conjugate operator then the role of $\Lambda$, \fer{202}, is taken by $N$, and relation \fer{seiv} is replaced by $\|P_{1,\beta} P(N\leq|\lambda|)\psi\|\geq 1-O(\lambda^0)$. But for $|\lambda|<1$, $P(N\leq |\lambda|)=|\Omega_{0}\rangle\langle \Omega_{0}|$ is just the projection the span of $\Omega_0$ (product of two vacua in $\ff\otimes\ff$), so one has $\psi=\Omega_{\beta,0}+O(\lambda^0)$, where $\Omega_{\beta,0}$ is the non-interacting KMS state.  Since $\Omega_{\beta,0}$ is close to $\Omega_{\beta,\lambda}^\xi$ for small values of $\lambda$, this means that there are no elements in the kernel of $\Llt$ which are orthogonal to $\Omega_{\beta,\lambda}^\xi$, provided $|\lambda|$ is small enough, i.e., the kernel of $\Llt$ has dimension one and consequently return to equilibrium holds. \\
\indent
The disadvantage of the translation generator is that its use requires (too) restricitve infrared regularity on the form factor. Indeed, the $j$-th commutator of the interaction with the translation generator involves the $j$-th derivative of the fuction $\frac{g}{\sqrt{e^{\beta\omega}-1}}$, so an infrared singular behaviour of this function is worsened by each application of the commutator (and we require those derivatives to be square integrable!). As a result, the case $g(0)\neq 0$ cannot be treated. \\
\indent
The remedy is to develop the theory with a conjugate operator $A$ which does not affect the infrared behaviour of $\frac{g}{\sqrt{e^{\beta\omega}-1}}$ in a negative way. The choice \fer{202.2} ({\it dilation generator}) is a good candidate (one could as well take operators interpolating between the translation and the dilation generator). The disadvantage of using the dilation generator is that its commutator with the non-interacting Liouvillian gives the operator $\Lambda$, which still has a one-dimensional kernel, but does {\it not} have a spectral gap at zero. This prevents us from showing that the kernel of $\Llt$ is simple. We can only prove \fer{seiv} which allows us only to show stability of $\ocond_{\beta,0}$, in the sense of Theorem \ref{stabthm}, but not return to equilibrium. \\
\indent
We remark that the dilation generator has been used in [BFSS] to show instability of excited eigenvalues in zero-temperature models. We expect that it is a relatively easy exercise to modify the techniques of [M1] and show absence of {\it nonzero} eigenvalues of $\Llt$ (which we view as the ``excited eigenvalues'' in the positive temperature case) by using the dilation instead of the translation generator. Notice though that if one succeeds to show that the kernel of $\Llt$ is simple, then one knows {\it automatically} that $\Llt$ cannot have any non-zero eigenvalues, see e.g. [JP2].

\subsubsection{Proof of Theorem \ref{stabthm}}

Fix $\eta>0, \xi\in[\rhocrit,\infty)\times S^1, B\in\frak A$ and choose an element $b'_{\xi,\eta}\in\pi_\xi({\frak A})'$ s.t. $\pi_\xi(B)\Omega_{\beta,0}-b'_{\xi,\eta}\Omega_{\beta,0} = O(\eta)$ ($\Omega_{\beta,0}$ is cyclic for the commutant $\pi_\xi({\frak A})'$). It follows that $\pi_\xi(B)\Omega^\xi_{\beta,\lambda}-b'_{\xi,\eta}\Omega_{\beta,0} = O(\eta +\|B\|\lambda^0)$ (we use that $\Omega^\xi_{\beta,\lambda}-\Omega^\xi_{\beta,0}=O(\lambda^0)$, see [FM2]) and consequently
\begin{eqnarray}
\lefteqn{
\scalprod{\pi_\xi(B)\Omega_{\beta,\lambda}^\xi}{e^{itL_{\lambda,\xi}}\pi_\xi(A)e^{-itL_{\lambda,\xi}}\pi_\xi(B)\Omega_{\beta,\lambda}^\xi}}\label{p1}\\
&=&\!\!\!\!\scalprod{\pi_\xi(B)\Omega_{\beta,\lambda}^\xi}{b'_{\xi,\eta}e^{itL_{\lambda,\xi}}\pi_\xi(A)\Omega_{\beta,\lambda}^\xi} +O\Big(\|A\|\,\|B\|\big(\eta + (\|B\|+\|b'_{\xi,\eta}\|)\lambda^0\big)\Big).
\nonumber
\end{eqnarray}
We use that $b'_{\xi,\eta}$ commutes with $e^{itL_{\lambda,\xi}}\pi_\xi(A)e^{-itL_{\lambda,\xi}}$, that $L_{\lambda,\xi}\Omega_{\beta,\lambda}^\xi=0$ and the above estimates. \\
\indent
The von Neumann ergodic theorem tells us that the ergodic average, $\frac 1T\int_0^T$, of the first term on the r.h.s. of \fer{p1} converges in the limit $T\rightarrow\infty$ to 
\begin{multline}
\scalprod{\pi_\xi(B)\Omega_{\beta,\lambda}^\xi}{b'_{\xi,\eta}\Pi_{\lambda,\xi} \pi_\xi(A)\Omega_{\beta,\lambda}^\xi}
=\scalprod{\pi_\xi(B)\Omega_{\beta,\lambda}^\xi}{b'_{\xi,\eta}\Omega_{\beta,\lambda}^\xi}\omega_{\beta,\lambda}^\xi(A)\\
+\sum_{j=1}^\infty \scalprod{\pi_\xi(B)\Omega_{\beta,\lambda}^\xi}{b'_{\xi,\eta}\psi_{j,\lambda}^\xi} \scalprod{\psi_{j,\lambda}^\xi}{\pi_\xi(A)\Omega_{\beta,\lambda}^\xi},
\label{p2}
\end{multline}
where $\Pi_{\lambda,\xi}$ is the projection onto the kernel of $L_{\lambda,\xi}$ and $\Omega_{\beta,\lambda}^\xi\cup\{\psi_{j,\lambda}^\xi\}_{j\geq 1}$ is an orthonormal basis of $\ker L_{\lambda,\xi}$. It follows from Corollary 2.4 that the series in \fer{p2} converges to zero as $\lambda\rightarrow 0$. Moreover we have 
\begin{equation}
\scalprod{\pi_\xi(B)\Omega_{\beta,\lambda}^\xi}{b'_{\xi,\eta}\Omega_{\beta,\lambda}^\xi}= \omega_{\beta,\lambda}^\xi(B^*B) +O\Big(\|B\|\big( \eta +(\|B\|+\|b'_{\xi,\eta}\|)\lambda^0\big)\Big),
\label{p3}
\end{equation}
where we use the estimates given at the beginning of the proof. The combination of \fer{p1}, \fer{p2}, \fer{p3} implies that there exists a $\lambda_1(\xi,\eta)>0$ s.t. if $0<|\lambda|<\lambda_1(\xi,\eta)$ then 
\begin{multline}
\lim_{T\rightarrow\infty} \frac 1T\int_0^T 
\scalprod{\pi_\xi(B)\Omega_{\beta,\lambda}^\xi}{e^{itL_{\lambda,\xi}}\pi_\xi(A)e^{-itL_{\lambda,\xi}}\pi_\xi(B)\Omega_{\beta,\lambda}^\xi}=\\
\left[\omega_{\beta,\lambda}^\xi(B^*B)+R_1\right]\omega_{\beta,\lambda}^\xi(A) +R_2,
\label{p4}
\end{multline}
where 
\begin{eqnarray*}
R_1&=& O\Big(\|B\|\big( \eta +(\|B\|+\|b'_{\xi,\eta}\|)\lambda^0\big)\Big),\\
R_2&=& O\Big( \|A\|\, \|B\|\big( \eta +(\|B\|+\|b'_{\xi,\eta}\|)\lambda^0\big)\Big).
\end{eqnarray*}
Given $\epsilon>0$ (as in the theorem) we can choose first $\eta$ small and then $\lambda$ small, in such a way that $R_1<\epsilon/2$ and $R_2<\|A\|\epsilon/2$.  The l.h.s. of \fer{p4} minus $\omega_{\beta,\lambda}^\xi(B^*B)\omega_{\beta,\lambda}^\xi(A)$ is then bounded in absolute value from above by  $\epsilon \|A\|$.
 \hfill $\blacksquare$

\section{Another abstract Virial Theorem with concrete applications}
\label{vtsection}

In this section we introduce a virial theorem in an abstract setting covering the cases of interest in the present paper (but which is general enough to allow for future generalizations). 
The virial theorem developed in [FM1], where the dominant part of $[L,A]$ commutes with $A$, does not apply to the present situation; here the leading term of $[[L,A],A]$ is $L$.

\subsection{The abstract Virial Theorem}
\label{substavt}

Let $\h$ be a Hilbert space, $\dom\subset\h$ a core for a selfadjoint
operator $Y\geq\bbbone$, and $X$ a symmetric operator on 
$\dom$. We say the triple $(X,Y,\dom)$ satisfies the {\it GJN
  (Glimm-Jaffe-Nelson) 
  Condition}, or that $(X,Y,\dom)$ is a {\it GJN-triple}, if there is a
constant $k<\infty$, s.t. for all $\psi\in\dom$: 
\begin{eqnarray}
\|X\psi\|&\leq& k\|Y\psi\| \label{nc1}\\
\pm i\left\{\scalprod{X\psi}{Y\psi}-\scalprod{Y\psi}{X\psi}\right\}&\leq&
k\scalprod{\psi}{Y\psi}.
\label{nc2}
\end{eqnarray}
Notice that if $(X_1,Y,\dom)$ and $(X_2,Y,\dom)$ are GJN triples, then so is
$(X_1+X_2,Y,\dom)$. Since $Y\geq\bbbone$, inequality \fer{nc1} is equivalent
to  
\begin{equation*}
\| X\psi\|\leq k_1\|Y\psi\|+k_2\|\psi\|,
\end{equation*}
for some $k_1, k_2<\infty$. Condition \fer{nc1} is phrased equivalently as ``$X\leq
kY$, in the sense of Kato on $\dom$''. 

\begin{theorem}{\bf (GJN commutator theorem) }
\label{nelsonthm}
\stepcounter{proposition}
If $(X,Y,\dom)$ satisfies
  the GJN Condition, then $X$ determines a selfadjoint operator (again
  denoted by $X$), s.t. $\dom(X)\supset\dom(Y)$. Moreover, $X$ is essentially
  selfadjoint on any core for $Y$, and \fer{nc1} is valid for all
  $\psi\in\dom(Y)$.
\end{theorem}

Based on the GJN commutator theorem we next
describe the setting for our general {\it virial theorem}. Suppose one is given
a selfadjoint operator $Y\geq\bbbone$ with core $\dom\subset\h$, and 
 operators  $L, A, \Lambda\geq 0, D, C_n$, $n=0,\ldots,4$, all symmetric on $\dom$, and being interrelated as 
\begin{eqnarray}
\scalprod{\varphi}{D\psi}&=&i\left\{
  \scalprod{L\varphi}{\Lambda\psi}-\scalprod{\Lambda\varphi}{L\psi}\right\} \label{44}\\
C_0&=& L\nonumber\\
\scalprod{\varphi}{C_n\psi}&=&i\left\{\scalprod{C_{n-1}\varphi}{A\psi}-\scalprod{A\varphi}{C_{n-1}\psi}\right\},\
  \ n=1,\ldots, 4,
\label{45}
\end{eqnarray}
where $\varphi, \psi\in\dom$. We assume that
\begin{itemize}
\item [(VT1)] $(X,Y,\dom)$ satisfies the GJN Condition, for
$X=L, \Lambda,D,C_n$. Consequently, all these operators determine selfadjoint
operators (which we denote by the same letters).
\item [(VT2)] $A$ is selfadjoint, $\dom\subset\dom(A)$, 
$e^{itA}$ leaves $\dom(Y)$ invariant, and 
\begin{equation}
e^{itA}Y e^{-itA} \leq k e^{k'|t|}Y,\ \ \ t\in\r,
\label{cond4}
\end{equation}
in the sense of Kato on $\dom$, for some constants $k, k'$.
\item [(VT3)] The operator $D$ satisfies $D\leq k\Lambda^{1/2}$ in the sense of Kato on $\dom$, for some constant $k$.
\item [(VT4)] Let the operators $V_n$ be defined as follows: for $n=1,3$ set $C_n= \Lambda +V_n$, and set $C_2=L_2+V_2$, $C_4=L_4+V_4$. We assume the following relative bounds, all understood in the sense of Kato on $\dom$:
\begin{eqnarray}
V_n &\leq& k\Lambda^{1/2}, \mbox{\ \ \ for $n=1,\ldots,4$},
\label{con1}\\
L_4 &\leq& k\Lambda,
\label{con2}\\
L_2 &\leq& k\Lambda^r, \mbox{\ \ \ for some $r>0$}.
\label{con3}
\end{eqnarray}
\end{itemize}
{\it Remark.\ }
The invariance condition $e^{itA}\dom(Y)\subset\dom(Y)$ implies that the bound \fer{cond4} holds in the sense of Kato on $\dom(Y)$, see [ABG], Propositions 3.2.2 and 3.2.5.

\begin{theorem}{\bf (Virial Theorem) }
\label{virialthm}
We assume the setting and assumptions introduced in this section so far. 
If $\psi\in\h$ is an eigenvector of $L$ then $\psi$ is in the form domain of $C_1$ and 
\begin{equation}
\av{C_1}_\psi=0.
\label{42}
\end{equation}
\end{theorem}
We prove this theorem in Section \ref{sectproofvthm}.

\subsection{The concrete applications}
\label{tca}

The proofs of Theorems \ref{vthm} and \ref{vthmnocond} reduce to an identification of the involved operators and domains and a subsequent verification of the assumptions of Section \ref{substavt}. Let us define
\begin{equation}
\dom = \cx^d\otimes\cx^d\otimes \ff_0(C^\infty_0(\r^3,d^3k))\otimes  \ff_0(C^\infty_0(\r^3,d^3k)),
\label{ver1}
\end{equation}
where $\ff_0$ is the finite-particle subspace of Fock space. Take
\begin{equation}
Y=\d\Gamma(\omega +1)\otimes\bbbone_\ff +\bbbone_\ff\otimes\d\Gamma(\omega+1) +\bbbone,
\label{ver2}
\end{equation}
and let the operators $L,\Lambda, A$ of Section \ref{substavt} be given, repectively,  by the operators $\Llt$ (see \fer{151.2}, or $L_\lambda$ in the case of Theorem \ref{vthmnocond}), \fer{202}, and \fer{202.2}. \\
\indent
It is an easy task to calculate the operators $C_j$; $C_1$ is given in \fer{202.1}, $C_2=L_2 +\lambda I_2$, $C_3=\Lambda +\lambda I_3$, $C_4=L_2+\lambda I_4$, where $L_2$ is given in \fer{129}, and where the $I_j$ are obtained similarly to $I_1$ (see \fer{202.4}). The operator $D$, \fer{44}, is just $i\lambda [I,\Lambda]$. It is a routine job to verify that Conditions (VT1)--(VT4) hold, with $V_n=I_n$ and $L_4=L_2$, $r=1$. 
To check Condition (VT2) one can use the explicit action of $e^{itA}$, see also [FM1], Section 8.

\section{Proof of Theorem \ref{virialthm}}
\label{sectproofvthm}

Before immersing ourselves into the details of the proof we present some facts we shall use repeatedly.
\begin{itemize}
\item If a unitary group $e^{itX}$ leaves the domain $\dom(Y)$ invariant then there exist constants $k, k'$ s.t. $\|Ye^{itX}\psi\|\leq ke^{k'|t|}\|Y\psi\|$, for all $\psi\in\dom(Y)$. Moreover, if $(X,Y,\dom)$ is a GJN triple then the unitary group $e^{itX}$ leaves $\dom(Y)$ invariant.
\item Let $(X,Y,\dom)$ and $(Z,Y,\dom)$ be GNS triples, and suppose that the quadratic form of the commutator of $X$ with $Z$, multiplied by $i$, is represented by a symmetric operator on $\dom$, denoted by $i[X,Z]$, and that $(i[X,Z], Y,\dom)$ is a GNJ triple. Then we have 
\begin{equation}
e^{itX} Z e^{-itX}-Z= \int_0^t dt_1 \,e^{it_1X}i[X,Z]e^{-it_1X}.
\label{77}
\end{equation}
This equality is understood in the sense of operators on $\dom(Y)$. Of course, if the higher commutators of $X$ with $Z$ also form GJN triples with $Y,\dom$ then one can iterate formula \fer{77}.
\end{itemize}
We refer to [FM1] and the references therein for more detail and further results of this sort. 
Let us introduce the cutoff functions
\begin{eqnarray}
f_1(x)&=&\int_{-\infty}^x dy\, e^{-y^2}, \ \ f(x)=e^{-y^2/2},\\
\label{51}
g&=&g_1^2,\label{50}
\end{eqnarray}
where $g_1\in C_0^\infty( (-1,1))$ satisfies $g_1(0)=1$. 
The derivative $(f_1)'$ equals $f^2$ which is strictly positive and the ratio $(f')^2/f$ decays faster than eponentially at infinity. 
The Gaussian $f$ is the fixed point of the Fourier transform
\begin{equation}
\widehat f(s)=(2\pi)^{-1/2}\int_\r dx\, e^{-isx} f(x),
\end{equation}
i.e., $\widehat f(s)=e^{-s^2/2}$, and we have $\widehat{(f_1)'\,}=is\widehat{f_1}=\widehat{f^2}$ which is a Gaussian itself. This means that $\widehat{f_1}$ decays like a Gaussian for large $|s|$ and has a singularity of type $s^{-1}$ at the origin. We define cutoff operators, for $\nu, \alpha>0$, by
\begin{eqnarray}
g_{1,\nu}&=&g_1(\nu\Lambda)=(2\pi)^{-1/2}\int_\r ds\, \widehat{g_1}(s)e^{is\nu\Lambda}\label{53}\\
g_\nu &=&g^2_{1,\nu}\label{54}\\
f_\alpha&=& f(\alpha A) =(2\pi)^{-1/2}\int_\r ds \widehat f(s) e^{is\alpha A}\label{55}.
\end{eqnarray}
Since $\widehat{f_1}$ has a singularity at the origin, we cut a small interval $(-\eta,\eta)$ out of the real axis, where $\eta>0$, and define
\begin{equation}
f_{1,\alpha}^\eta =\alpha^{-1} (2\pi)^{-1/2}\int_\reta ds \,\widehat{f_1}(s) e^{is\alpha A},
\label{56}
\end{equation}
where we set $\reta=\r\backslash (-\eta,\eta)$. Standard results about invariance of domains show that the cutoff operators $g_\nu, f_\alpha, f_{1,\alpha}^\eta$ are bounded selfadjoint operators leaving the domain $\dom(Y)$ invariant, and it is not hard to see that $\|f_{1,\alpha}^\eta\|\leq k/\alpha$, uniformly in $\eta$ (see [FM1]). \\
\indent
Suppose that $\psi$ is a normalized eigenvector of $L$ with eigenvalue $e$, $L\psi=e\psi$, $\|\psi\|=1$. Let $\varphi\in\h$ be s.t. $\psi=(L+i)^{-1}\varphi$ and let $\{\varphi_n\}\subset\dom$ be a sequence approximating $\varphi$, $\varphi_n\rightarrow\varphi$. Then we have 
\begin{equation}
\psi_n=(L+i)^{-1}\varphi_n\longrightarrow\psi,\ \ n\rightarrow\infty, 
\label{57}
\end{equation}
and $\psi_n\in\dom(Y)$. The latter statement holds since the resolvent of $L$ leaves $\dom(Y)$ invariant (which in turn is true since $(L,Y,\dom)$ is a GJN triple). It follows that the {\it regularized eigenfunction}
\begin{equation}
\psi_{\alpha,\nu,n}=f_\alpha g_\nu \psi_n
\label{58}
\end{equation}
is in $\dom(Y)$, and that $\psi_{\alpha,\nu,n}\rightarrow\psi$, as $\alpha,\nu\rightarrow 0$ and $n\rightarrow\infty$. It is not hard to see that $(L-e)\psi_n\rightarrow 0$ as $n\rightarrow\infty$, a fact we write as
\begin{equation}
(L-e)\psi_n=o(n).
\label{59}
\end{equation}
Since $f_{1,\alpha}^\eta$ leaves $\dom(Y)$ invariant, and since $\dom(Y)\subset\dom(L)$, the commutator $-i[f_{1,\alpha}^\eta,L]$ is defined in the usual (strong) way on $\dom(Y)$. We consider its expectation value in the state $g_\nu\psi_n\in\dom(Y)$,
\begin{equation}
-i\av{[f_{1,\alpha}^\eta,L]}_{g_\nu\psi_n}=-i\av{[f_{1,\alpha}^\eta,L-e]}_{g_\nu\psi_n}. 
\label{60}
\end{equation}
The idea is to write \fer{60} on the one hand as $\av{C_1}_{\psiann}$ modulo some small term for appropriate $\alpha,\nu,n$ (``positive commutator''), and on the other hand to see that \fer{60} itself is small, using the fact that $(L-e)\psi=0$. \\
\indent
The latter is easily seen by first writing
\begin{equation}
(L-e)g_\nu\psi_n= g_\nu (L-e)\psi_n +g_{1,\nu}[L,g_{1,\nu}]\psi_n +[L,g_{1,\nu}]g_{1,\nu}\psi_n
\label{75}
\end{equation}
and then realizing that, due to condition (VT3),
\begin{equation*}
g_{1,\nu}[L,g_{1,\nu}] =\frac{\nu}{(2\pi)^{1/2}}\int_\r ds\, \widehat{g_1}(s) e^{is\nu \Lambda}\int_0^s ds_1\, e^{-is_1\nu \Lambda} g_{1,\nu} D e^{is\nu\Lambda} =\O{\sqrt\nu},
\end{equation*}
and similarly, $[L,g_{1,\nu}]g_{1,\nu}=\O{\sqrt\nu}$, so that 
\begin{equation}
-i\av{[f_{1,\alpha}^\eta, L]}_{g_\nu\psi_n} =\O{\frac{o(n)+\sqrt\nu}{\alpha}}.
\label{76}
\end{equation}
\indent
Next we figure out a lower bound on \fer{60}. A repeated application of formula \fer{77} gives, in the strong sense on $\dom(Y)$,
\begin{eqnarray}
\lefteqn{
-i[f_{1,\alpha}^\eta,L]= f_{1,\alpha}'C_1 -i\frac{\alpha}{2!} f_{1,\alpha}'' C_2 -\frac{\alpha^2}{3!}f'''_{1,\alpha}C_3} \nonumber\\
&&+\frac{i\alpha^3}{(2\pi)^{1/2}}\int_\reta ds\, \widehat{f_1}(s) e^{is\alpha A}\int_0^sds_1 \int_0^{s_1}ds_2\int_0^{s_2}ds_3\int_0^{s_4} ds_4\, e^{-is_4\alpha A} C_4 e^{is_4\alpha A}\nonumber\\
&&+ R_{\eta,1} C_1 +\frac{\alpha}{2!} R_{\eta,2}C_2 +\frac{\alpha^2}{3!} C_3,
\label{61}
\end{eqnarray}
where we use that 
\begin{equation*}
(2\pi)^{-1/2}\int_\r ds\, (is)^n\widehat f(s) e^{isx} =f^{(n)}(x),
\end{equation*}
and where we set $f'_{1,\alpha}=(f_1)'(\alpha A)$, $f''_{1,\alpha}=(f_1)''(\alpha A)$, e.t.c., and 
\begin{equation}
R_{\eta,n}=-i (2\pi)^{-1/2}\int_{-\eta}^\eta ds\, s^n \widehat{f_1}(s) e^{is\alpha A}.
\label{63}
\end{equation}
Using that $f'_{1,\alpha}=f^2(\alpha A)=f_\alpha^2$ and applying again expansion \fer{77} yields
\begin{eqnarray}
\lefteqn{
f'_{1,\alpha} C_1=f_\alpha C_1 f_\alpha +i\alpha f_\alpha f'_\alpha C_2 +\frac{\alpha^2}{2!} f_\alpha f''_\alpha C_3
\label{64''}}\\
&&- \frac{\alpha^3}{(2\pi)^{1/2}} f_\alpha\int_\r ds\, \widehat f(s) e^{is\alpha A}\int_0^sds_1 \int_0^{s_1}ds_2\int_0^{s_2} ds_3\, e^{-is_3 \alpha A} C_3 e^{is_3\alpha A}.
\nonumber
\end{eqnarray}
Plugging this into the r.h.s. of \fer{61} and using that $f''_{1,\alpha}=2f_\alpha f'_\alpha$, we obtain
\begin{eqnarray}
\lefteqn{
-i\av{[f_{1,\alpha}^\eta,L]}_{g_\nu\psi_n}} \label{64'}    \\
&=&\av{C_1}_{\psiann} +\alpha^2 \RE \av{\frac{1}{2} f_\alpha f''_\alpha C_3 -\frac{1}{3!} f'''_{1,\alpha} C_3}_{g_\nu\psi_n} 
+ \O{\frac{\eta}{\nu^r} +\frac{\eta}{\sqrt{\nu}}+\frac{\alpha^3}{\nu}}.
\nonumber
\end{eqnarray}
We take the real part on the r.h.s. for free since the l.h.s. is real. 
The error term in \fer{64'} is obtained as follows. Clearly we have $R_{\eta,n}=\O{\eta}$ and condition (VT4) gives $C_ng_\nu=\O{\nu^{-r}+\nu^{-1/2}}$, which accounts for the term $\O{\eta/\nu^r+\eta/\sqrt{\nu}}$. The term $\O{\alpha^3/\nu}$ is an upper bound for the expectation of the terms in \fer{61} and \fer{64''} involving the multiple integrals, in the state $g_\nu\psi_n$. For instance, the contribution coming from \fer{61} is bounded above as follows. Due to condition (VT4) we have
\begin{equation*}
\|e^{-is_4 \alpha  A}C_4 e^{is_4\alpha A}g_\nu\psi_n\|\leq k \|\Lambda e^{is_4\alpha A} g_\nu\psi_n\| =e^{k'\alpha |s_4|} \O{\frac 1\nu}, 
\end{equation*}
which gives the following upper bound on the relevant term: 
\begin{equation*}
\alpha^3\int_\reta ds\, \left| \widehat{f_1}(s)\right| s^4 e^{k'|s|} \cdot \O{\frac 1\nu}.
\end{equation*}
The integral is finite because $\widehat{f_1}$ has Gaussian decay. \\
\indent 
Our next task is to esimtate the real part in \fer{64'}. It suffices to consider the terms
\begin{equation}
\alpha^2 \RE\av{f''_\alpha f_\alpha C_3}_{g_\nu\psi_n}\mbox{\ \ \ and\ \ \ \ } \alpha^2 \RE\av{(f_\alpha')^2 C_3}_{g_\nu\psi_n},
\label{80}
\end{equation}
because $f'''_{1,\alpha}=2(f'_\alpha)^2 +2 f''_\alpha f_\alpha$. Let us start with the first term in \fer{80}. Using the decompostion $C_3=\Lambda +V_3$ and the relative bound of $V_3$ given in (VT4) we estimate
\begin{eqnarray}
\alpha^2\RE\av{f''_\alpha f_\alpha C_3}_{g_\nu\psi_n} &=& \alpha^2 \RE\av{f''_\alpha f_\alpha \Lambda}_{g_\nu\psi_n} +\O{\frac{\alpha^2}{\sqrt{\nu}}}\nonumber\\
&=& \alpha^2 \RE\av{f''_\alpha \Lambda f_\alpha}_{g_\nu\psi_n} +\O{\frac{\alpha^2}{\sqrt \nu}+\frac{\alpha^3}{\nu}}.
\label{66}
\end{eqnarray}
We bound the first term on the r.h.s. from above as
\begin{equation}
\alpha^2  \left|\RE\av{f''_\alpha \Lambda f_\alpha}_{g_\nu\psi_n}\right| \leq \alpha^2 \|\Lambda^{1/2}f''_\alpha g_\nu\psi_n\|\, \|\Lambda^{1/2}\psiann\|
\label{66'}
\end{equation}
and use that 
\begin{equation*}
\av{f''_\alpha \Lambda f''_\alpha}_{g_\nu\psi_n}\leq \int_\r ds\, |\widehat{f''}(s)| \, \left|\av{f''_\alpha \Lambda e^{is\alpha A}}_{g_\nu\psi_n}\right|
=\O{\frac 1\nu}
\end{equation*}
 to see that for any $c>0$, 
\begin{equation}
\alpha^2 \left| \RE\av{f''_\alpha \Lambda f_\alpha}_{g_\nu\psi_n}\right|  \leq \frac{\alpha^4}{c \nu}+c\av{\Lambda}_\psiann.
\label{67}
\end{equation}
Choose $c=\alpha^{1+\xi}$, for some $\xi>0$ to be determined later. Then, inserting again a term $V_1$ into the last expectation value (by adding a correction of size $\O{\alpha^{1+\xi}/\sqrt\nu}$), we get 
\begin{equation}
\left|\fer{66}\right|\leq  \alpha^{1+\xi}\left|\av{C_1}_\psiann\right| +\O{\frac{\alpha^2}{\sqrt\nu} +\frac{\alpha^3}{\nu}+\frac{\alpha^{1+\xi}}{\sqrt\nu} +\frac{\alpha^{3-\xi}}{\nu}}.
\label{68}
\end{equation}
Next we tackle the second term in \fer{80}. The Gaussian $f$ is strictly positive, so we can write
\begin{eqnarray}
\alpha^2\RE\av{(f_\alpha')^2C_3}_{g_\nu\psi_n} &=&
\alpha^2 \RE\av{\frac{(f_\alpha')^2}{f_\alpha}f_\alpha C_3}_{g_\nu\psi_n} \nonumber\\
&=&\alpha^2 \RE\av{\frac{(f_\alpha')^2}{f_\alpha}\Lambda f_\alpha}_{g_\nu\psi_n} +\O{\frac{\alpha^2}{\sqrt\nu}+\frac{\alpha^3}{\nu}},\ \ \ 
\label{69}
\end{eqnarray}
where we have taken into account condition (VT4) in the same way as above. It follows that 
\begin{equation*}
\left|\fer{69}\right| \leq \alpha^2 \left\|\Lambda^{1/2}\frac{(f_\alpha')^2}{f_\alpha}g_\nu\psi_n\right\|\, \|\Lambda^{1/2} \psiann\| +\O{\frac{\alpha^2}{\sqrt\nu}+\frac{\alpha^3}{\nu}},
\end{equation*}
and proceeding as in \fer{66'}--\fer{67} we see that 
\begin{equation}
\alpha^2\left| \av{(f_\alpha')^2 C_3}_{g_\nu\psi_n}\right|
\leq \alpha^{1+\xi}\left| \av{C_1}_\psiann\right|  +\O{\frac{\alpha^2}{\sqrt\nu} +\frac{\alpha^3}{\nu}+\frac{\alpha^{1+\xi}}{\sqrt\nu} +\frac{\alpha^{3-\xi}}{\nu}}.
\label{70}
\end{equation}
Estimates \fer{68} and \fer{70} together with \fer{64'} give the bound
\begin{eqnarray}
\lefteqn{
\left|-i\av{[f_{1,\alpha}^\eta,L]}_{g_\nu\psi_n}\right| \geq \left(1-\O{\alpha^{1+\xi}}\right)\left|\av{C_1}_\psiann\right|}\nonumber \\
&& +\O{\frac{\alpha^2}{\sqrt\nu} +\frac{\alpha^3}{\nu}+\frac{\alpha^{1+\xi}}{\sqrt\nu} +\frac{\alpha^{3-\xi}}{\nu}+\frac{\eta}{\nu^r}+\frac{\eta}{\sqrt{\nu}}}.\label{72}
\end{eqnarray}
We combine this upper bound with the lower bound obtained in \fer{76} to arrive at
\begin{eqnarray}
\lefteqn{
\left(1-\O{\alpha^{1+\xi}}\right)\left|\av{C_1}_\psiann\right|}\label{74}\\
&& =
\O{\frac{\sqrt{\nu} +o(n)}{\alpha} +\frac{\alpha^2}{\sqrt\nu} +\frac{\alpha^3}{\nu}+\frac{\alpha^{1+\xi}}{\sqrt\nu} +\frac{\alpha^{3-\xi}}{\nu}+\frac{\eta}{\nu^r}+\frac{\eta}{\nu}}. 
\nonumber
\end{eqnarray}
Choose $\alpha$ so small that $1-\O{\alpha^{1+\xi}}>1/2$ and take the limits $\eta\rightarrow 0$, $n\rightarrow\infty$ to get 
\begin{equation}
\left|\av{C_1}_{f_\alpha g_\nu\psi}\right|
=\O{\frac{\sqrt{\nu}}{\alpha} +\frac{\alpha^2}{\sqrt\nu} +\frac{\alpha^3}{\nu}+\frac{\alpha^{1+\xi}}{\sqrt\nu} +\frac{\alpha^{3-\xi}}{\nu}}. 
\label{74a}
\end{equation}
Take for example $\xi=1/2$, $\nu=\nu(\alpha)=\alpha^{9/4}$. Then the r.h.s. of \fer{74a} is $\O{\alpha^{1/4}}$, so
\begin{equation*}
\lim_{\alpha\rightarrow 0}\av{C_1}_{f_\alpha g_{\nu(\alpha)}\psi}=0.
\end{equation*}
Since the operator $C_1$ is semibounded its quadratic form is closed, hence it follows from $f_\alpha g_{\nu(\alpha)}\psi\rightarrow \psi$, $\alpha\rightarrow 0$, that $\psi$ is in the form domain of $C_1$ and that $\av{C_1}_\psi=0$.\hfill $\blacksquare$

\section{Proofs of Theorem \ref{chev} and of Corollary 2.4}
\label{ptchev}

In order to alleviate the notation we drop in this section the variable $\xi$ labelling the fiber in the decomposition \fer{135'} (imagining $\xi\in[\rhocrit,\infty)\times S^1$ to be fixed). The operator $\Llt$, \fer{151.2}, is thus denoted
\begin{equation}
L_\lambda =L_0 +\lambda (I +K),
\label{201}
\end{equation}
where $I$ and $K$ are given in \fer{153.1}, \fer{153.153}. 
In parallel we can imagine that $K=0$ and that Condition (A1) is replaced by \fer{n1}. \\
\indent
Let $\epsilon,\rho,\theta>0$ be parameters. Set
\begin{eqnarray}
P_\rho &=& P_0 P(\Lambda\leq \rho) \label{203}\\
P_0&=& P(L_1=0)\nonumber\\
A_0&=& i\theta\lambda (P_\rho I \repsilonbar^2 -\repsilonbar^2 I P_\rho)\label{204}\\
\repsilonbar&=& \Pbar_\rho R_\epsilon\nonumber\\
R_\epsilon&=& (L^2_0+\epsilon^2)^{-1/2} \label{205}
\end{eqnarray}
where $\Pbar_\rho=\bbbone-P_\rho$. We  also set $\Pbar_0=\bbbone-P_0$. The product in \fer{203} is understood in the sprit of leaving out trivial factors ($P_\rho = P_0\otimes P(\Lambda\leq\rho)$). We also define the selfadjoint operator (c.f. \fer{202.1}, \fer{202.4})
\begin{equation}
B= C_1+i[L_\lambda,A_0]=\Lambda +I_1 +i [L_\lambda,A_0],
\label{206}
\end{equation}
where the last commutator is a bounded operator. Let us decompose
\begin{equation}
B=P_\rho BP_\rho +\Pbar_\rho B\Pbar_\rho +2\RE P_\rho B\Pbar_\rho.
\label{207}
\end{equation}
Our goal is to obtain a lower bound on $\av{B}_{\psi_\lambda}$, the expectation value of $B$ in the state given by the normalized eigenvector $\psi_\lambda$ of $L_\lambda$. We look at each term in \fer{207} separately. 
In what follows we use the standard form bound 
\begin{equation}
\lambda I_1\geq -\frac 12\Lambda -\O{\lambda^2},
\label{standardformbound}
\end{equation}
and the estimates $\|\Lambda^{1/2}\psi_\lambda\|=\O{\lambda}$, $\| \Pbar_0 P(\Lambda\leq\rho)\psi_\lambda\|=\O{\lambda}$. The former estimate follows from Theorem \ref{vthm} (or Theorem \ref{vthmnocond} for the system without condensate) and the latter is easily obtained like this: let $\chi\in C_0^\infty(\r)$ be such that $0\leq\chi\leq 1$, $\chi(0)=1$ and such that $\chi$ has support in a neighborhood of the origin containing no other eigenvalue of $L_1$ than zero. Then, for $\rho$ sufficiently small, we have $\Pbar_0 P(\Lambda\leq\rho)\chi(L_0)=0$, so $\Pbar_0 P(\Lambda\leq\rho)\psi_\lambda =\Pbar_0 P(\Lambda\leq\rho)(\chi(L_\lambda)-\chi(L_0))\psi_\lambda =\O{\lambda}$, by standard functional calculus. \\
\indent
Taking into account \fer{standardformbound} we estimate
\begin{eqnarray}
\lefteqn{
\av{P_\rho BP_\rho}_{\psi_\lambda}}\nonumber\\
 &\geq& -\theta\lambda^2 \av{P_\rho [I+K,P_\rho I \repsilonbar^2  -\repsilonbar^2 IP_\rho] P_\rho}_{\psi_\lambda}-\O{\lambda^2}\nonumber\\
&=& 2\theta\lambda^2\av{P_\rho I \repsilonbar^2 IP_\rho}_{\psi_\lambda}
+\theta\lambda^2\av{P_\rho I \repsilonbar^2  K P_\rho +P_\rho K\repsilonbar^2 IP_\rho}_{\psi_\lambda} -\O{\lambda^2}\nonumber\\
&\geq& 2\theta\lambda^2\av{P_\rho I \repsilonbar^2 IP_\rho}_{\psi_\lambda}
-\frac{\theta\lambda^2}{\epsilon} \O{\frac{\epsilon}{\theta}+\epsilon},
\label{208}
\end{eqnarray}
where we use in the last step that $\Pbar_\rho=\Pbar_0 P(\Lambda\leq\rho) +P(\Lambda>\rho)$ to arrive at
\begin{equation*}
\| P_\rho I\repsilonbar^2 KP_\rho\|  =\| P_\rho I R_\epsilon^2 \Pbar_0 P(\Lambda\leq\rho) K P_\rho\|\leq c.
\end{equation*}
The last estimate is due to $\|R_\epsilon^2 \Pbar_0 P(\Lambda\leq\rho)\|<c$ and $\| P_\rho I P(\Lambda<\rho)\| < c$ (this follows in a standard way assuming condition \fer{n2}). \\
\indent
Next we estimate
\begin{equation}
\av{\Pbar_\rho B\Pbar_\rho}_{\psi_\lambda}
\geq \frac 12 \av{\Pbar_\rho \Lambda}_{\psi_\lambda} -2\theta\lambda^2\RE \av{\Pbar_\rho (I+K) P_\rho I\repsilonbar^2 }_{\psi_\lambda} -\O{\lambda^2}
\label{209}
\end{equation}
and 
\begin{equation}
\av{\Pbar_\rho (I+K) P_\rho I\repsilonbar^2 }_{\psi_\lambda}
= \|\Pbar_\rho\psi_\lambda\|^2\  \O{\frac{1}{\epsilon} \|P_\rho I\repsilon\|}
=\O{\frac{\lambda^2}{\rho\epsilon^{3/2}}},
\label{210}
\end{equation}
where we use $\|\Pbar_\rho\psi_\lambda\|\leq \|\Pbar_0 P(\Lambda\leq\rho)\psi_\lambda\| +\| P(\Lambda>\rho)\psi_\lambda\|=\O{\lambda/\sqrt\rho}$, and $\|P_\rho IR_\epsilon\|=\O{1/\sqrt\epsilon}$. The former estimate follows from the observations after \fer{standardformbound} and from $\|P(\Lambda>\rho)\psi_\lambda\|\leq 1/\sqrt\rho \|P(\Lambda>\rho)\Lambda^{1/2}\psi_\lambda\|=O(\lambda/\sqrt\rho)$. The estimate $\|P_\rho IR_\epsilon\|=\O{1/\sqrt\epsilon}$ is standard in this business, it follows from $P_\rho I R^2_\epsilon I P_\rho=\O{1/\epsilon}$ (see e.g. [BFSS] and also the explanations before \fer{216} here below). Combining \fer{209} and \fer{210}, and taking into account that $\av{\Pbar_\rho\Lambda}_{\psi_\lambda} \geq \av{P(\Lambda>\rho)\Lambda}_{\psi_\lambda}\geq \rho\av{P(\Lambda>\rho)}_{\psi_\lambda}\geq \rho(\av{\Pbar_\rho}_{\psi_\lambda} -\O{\lambda^2})$ gives
\begin{equation}
\av{\Pbar_\rho B\Pbar_\rho}_{\psi_\lambda}
\geq\frac\rho 2 \av{\Pbar_\rho}_{\psi_\lambda} -\frac{\theta\lambda^2}{\epsilon} \O{\frac\epsilon\theta +\frac{\lambda^2}{\rho\sqrt\epsilon}}.
\label{211}
\end{equation}
\indent
Our next task it to estimate
\begin{equation}
\av{P_\rho B\Pbar_\rho}_{\psi_\lambda} = \lambda\av{P_\rho I_1\Pbar_\rho}_{\psi_\lambda} -\theta\lambda\av{P_\rho (L_\lambda P_\rho IR_\epsilon^2 -I \repsilonbar^2 L_\lambda) \Pbar_\rho}_{\psi_\lambda}.
\label{212}
\end{equation}
It is not difficult to see that 
\begin{eqnarray*}
\av{P_\rho I_1\Pbar_\rho}_{\psi_\lambda} &=&\av{P_\rho I_1 \Pbar_0 P(\Lambda\leq\rho)}_{\psi_\lambda} + \av{ P_\rho I_1 P(\Lambda>\rho)}_{\psi_\lambda}\\
&=&\O{\lambda}+ \O{\|(I_1)_{\rm a}\Lambda^{-1/2}\|\, \|\Lambda^{1/2}\psi_\lambda\|}\\
&=&\O{\lambda},
\end{eqnarray*}
where $(I_1)_{\rm a}$ means that we take in $I_1$ only the terms containing annihilation operators (see \fer{153.1}) and where we use $\| (I_1)_{\rm a}\Lambda^{-1/2}\|<c$. The second term on the r.h.s. of \fer{212} is somewhat more difficult to estimate. We have
\begin{eqnarray}
\lefteqn{
\theta\lambda\av{P_\rho (L_\lambda P_\rho I R_\epsilon^2 -I \repsilonbar^2  L_\lambda) \Pbar_\rho}_{\psi_\lambda}}\nonumber\\
&=&-\theta\lambda^2\av{(I+K)P_\rho  I\repsilonbar^2}_{\psi_\lambda} -\theta\lambda\av{P_\rho IR_\epsilon^2 L_0\Pbar_\rho}_{\psi_\lambda}\nonumber\\
&&+\theta\lambda^2\av{P_\rho( (I+K)P_\rho I\repsilonbar^2 -I \repsilonbar^2 \Pbar_\rho (I+K))\Pbar_\rho}_{\psi_\lambda},
\label{213}
\end{eqnarray}
where the first term on the r.h.s. comes from the contribution $\av{P_\rho L_0 I\repsilonbar^2}_{\psi_\lambda}$ in the l.h.s. by using that $P_\rho L_0=L_0P_\rho=L_\lambda P_\rho-\lambda (I+K)P_\rho$ and that $L_\lambda\psi_\lambda=0$. We treat the first term on the r.h.s. of \fer{213} as
\begin{eqnarray}
\lefteqn{
\av{(I+K) P_\rho I \repsilonbar^2}_{\psi_\lambda}}\nonumber\\
&=& \av{(I+K) P_\rho I R_\epsilon^2 \Pbar_0 P(\Lambda\leq\rho)}_{\psi_\lambda} +\av{(I+K) P_\rho I R_\epsilon^2 P(\Lambda>\rho)}_{\psi_\lambda}\nonumber\\
&=& \O{\lambda \|\Pbar_0 P(\Lambda\leq\rho)\psi_\lambda\|}
+ \O{\epsilon^{-2} \|(I_1)_{\rm a}\Lambda^{-1/2}\|\, \|\Lambda^{1/2}\psi_\lambda\|}\nonumber\\
&=&\O{\lambda+\frac{\lambda}{\epsilon^2}}.
\label{214}
\end{eqnarray}
The second term on the r.h.s. of \fer{213} has the bound
\begin{eqnarray}
\av{P_\rho IR_\epsilon^2 L_0\Pbar_\rho}_{\psi_\lambda}
&=&\av{P_\rho I R_\epsilon^2 L_0 \Pbar_0  P(\Lambda\leq\rho)}_{\psi_\lambda} +\av{P_\rho I R_\epsilon^2 L_0 P(\Lambda>\rho)}_{\psi_\lambda}\nonumber\\
&=&\O{\lambda} +\O{ \|P_\rho (I)_{\rm a} R_\epsilon \Lambda^{-1/2} P(\Lambda>\rho)\|\, \|\Lambda^{1/2}\psi_\lambda\|}\nonumber\\
&=&\O{\frac{\lambda}{\sqrt\epsilon}},
\label{215}
\end{eqnarray}
where we use that (with $(I)_{\rm c}= ((I)_{\rm a})^*$) 
\begin{equation*}
\|P_\rho (I)_{\rm a} R_\epsilon \Lambda^{-1/2} P(\Lambda>\rho)\|^2 =
\| P(\Lambda>\rho) \Lambda^{-1/2} R_\epsilon (I)_{\rm c} P_\rho\|^2
=\O{\frac 1\epsilon}.
\end{equation*}
The latter bound can be shown by using the explicit form of the interaction $I$, given in \fer{153.1}, and by using standard pull-through formulae to see that a typical contraction term in  $P_\rho (I)_{\rm a} R_\epsilon^2 \Lambda^{-1} P(\Lambda>\rho) (I)_{\rm c}P_\rho$ has the form
\begin{equation*}
\int d^3k \ \frac{|g(k)|^2}{e^{\beta\omega}-1} P_\rho (G_\pm\otimes\bbbone_{\cx^d}) \frac{P(\Lambda+|k|>\rho)}{(\Lambda +|k|)\, ((L_0\pm\omega)^2+\epsilon^2)} (G_\pm\otimes\bbbone_{\cx^d}) P_\rho
\end{equation*}
and is thus bounded from above, in norm, by a constant times $1/\epsilon$, provided $p>-1/2$ (recall that $p$ characterizes the infrared behaviour of the form factor, see Theorem \ref{t'}; in the case of the system with condensate we have $p=0$).  To see this use $(\Lambda+|k|)^{-1}\leq |k|^{-1}$, and then standard estimates which show that the resulting operator is of order $\epsilon^{-1}$; the mechanism is that the main part comes from the restriction of the operator to $\ran P_0 P_{\Omega_0}$ ($\rho=0$) and there the resolvent, when multiplied by $\epsilon$, converges to the Dirac delta distribution $\delta(L_1\pm \omega)$, so the integral is $1/\epsilon$ times a bounded operator. See also Lemma 6.4 of [BFSS]. \\
\indent
Next we estimate the third term in the r.h.s. of \fer{213} as 
\begin{eqnarray}
\lefteqn{
\av{ P_\rho( (I+K) P_\rho I R_\epsilon^2 - I \repsilonbar^2  (I+K))\Pbar_\rho}_{\psi_\lambda}}\nonumber\\
&=&\O{\epsilon^{-3/2} \| \Pbar_\rho\psi_\lambda\|} +\O{\|P_\rho I \repsilonbar^2 (I+K)\Pbar_\rho \psi_\lambda\|}\nonumber\\
&=&\O{\frac{\lambda}{\sqrt\rho \epsilon^2}},
\label{216}
\end{eqnarray}
where we use again that $\|P_\rho I R_\epsilon\|=\O{1/\sqrt\epsilon}$, $\|\Pbar_\rho\psi_\lambda\|=O(\lambda/\sqrt\rho)$, and that $\| P_\rho I \repsilonbar^2 I\|=\O{1/\epsilon^2}$. Collecting the effort we put into estimates \fer{214}, \fer{215} and \fer{216} rewards us with the bound
\begin{equation}
\av{P_\rho B \Pbar_\rho}_{\psi_\lambda} = \frac{\theta\lambda^2}{\epsilon}\ \O{\frac\epsilon\theta +\sqrt\epsilon +\frac{\lambda}{\epsilon\sqrt\rho}},
\label{217}
\end{equation}
which we combine with \fer{208} and \fer{211} to obtain 
\begin{eqnarray}
\lefteqn{
\av{B}_{\psi_\lambda}}\label{218}\\
&\geq& 2\theta\lambda^2 \av{P_\rho I \repsilonbar^2  IP_\rho}_{\psi_\lambda} +\frac\rho 2\av{\Pbar_\rho}_{\psi_\lambda}-\frac{\theta\lambda^2}{\epsilon} \O{\frac\epsilon\theta +\frac{\lambda^2}{\rho\sqrt\epsilon}  +\sqrt\epsilon+\frac{\lambda}{\epsilon \sqrt\rho}}.
\nonumber
\end{eqnarray}
The non-negative operator $P_\rho I \repsilonbar^2 IP_\rho$ has appeared in various guises in many previous papers on the subject (``level shift operator'').  The following result follows from a rather straightforward calculation, using the explicit form of the interaction $I$, \fer{153.1}. We do not write down the analysis, one can follow closely e.g. [BFSS], [M1], [BFS].
\begin{lemma}
We have the expansion 
\begin{equation}
P_\rho I \repsilonbar^2 IP_\rho= \frac 1\epsilon P_0\big(\Gamma +O(\epsilon^0)\big)P_0\otimes P(\Lambda\leq \rho) +\O{\frac{\rho^{2+2p}}{\epsilon^2}+\frac{\rho}{\epsilon^3}},
\label{219}
\end{equation}
where $p$ is the parameter characterizing the infrared behaviour of the form factor (see Theorem \ref{t'}; in the situation of Theorem \ref{chev} we set $p=0$), $O(\epsilon^0)$ is an operator whose norm vanishes in the limit $\epsilon\rightarrow 0$, and where 
\begin{equation}
\Gamma =\widetilde\Gamma\ \int_{S^2}d\sigma|g(1,\sigma)|^2,
\label{r100}
\end{equation}
and $\widetilde \Gamma$ is the non-negative operator on $\ran P_0$ which has the following matrix representation in the basis $\{\varphi_0\otimes\varphi_0,\ldots,\varphi_{d-1}\otimes\varphi_{d-1}\}$ of $\ran P_0$: $\widetilde \Gamma$ is tridiagonal with diagonal given by $[a,1+2a,\ldots,1+2a,1+a]$ and constant subdiagonal and superdiagonal with entries $-\sqrt{a(1+a)}$,  $a=\rho(1)=\frac{1}{e^{\beta}-1}$. The kernel of $\Gamma$ is spanned by the Gibbs state \fer{gibbsvector}, $\ker(\Gamma)=\cx\Omega_{1,\beta}$, and the spectrum of $\Gamma$ has a gap $\gamma>0$ at zero which is uniform in $\beta\geq\beta_0$, for $\beta_0$ fixed.
\end{lemma}
{\it Remark.\ } It is easily verified by explicit calculation that $\Omega_{1,\beta}$ is the unique element in the kernel of $\widetilde\Gamma$. To see that the gap $\gamma$ is independent of $\beta$ for large $\beta$ one can use the fact that $\widetilde\Gamma$ converges to the matrix $\rm{diag}[0,1,1,\ldots,1]$ in the limit $\beta\rightarrow\infty$. The latter matrix has a gap of size one. \\
\indent
It follows from the lemma that 
\begin{eqnarray}
\lefteqn{
2\theta\lambda^2\av{P_\rho I \repsilonbar^2  IP_\rho}_{\psi_\lambda}}\nonumber\\
&\geq& 2\frac{\theta\lambda^2}{\epsilon}\gamma \av{\Pbar_{1,\beta}P_\rho}_{\psi_\lambda} -\frac{\theta\lambda^2}{\epsilon}\ \left(O(\epsilon^0)+ \O{\frac{\rho^{2+2p}}{\epsilon} +\frac{\rho}{\epsilon^2}}\right),\ \ 
\label{222}
\end{eqnarray}
where $\Pbar_{1,\beta}=\bbbone-P_{1,\beta}$, and $P_{1,\beta}=|\Omega_{1,\beta}\rangle\langle\Omega_{1,\beta}|$ is the projection onto the span of the Gibbs state \fer{gibbsvector}. Using this estimate in \fer{218} gives
\begin{eqnarray}
\av{B}_{\psi_\lambda} &\geq& \min\left\{\frac{2\theta\lambda^2}{\epsilon}\gamma, \frac{\rho}{2}\right\} \|\psi_\lambda\|^2 -\frac{2\theta\lambda^2}{\epsilon}\gamma\av{P_{1,\beta}P(\Lambda\leq\rho)}_{\psi_\lambda}\nonumber\\
&&- \frac{\theta\lambda^2}{\epsilon}\ \O{\frac\epsilon\theta +\frac{\lambda^2}{\rho\sqrt\epsilon}  +\frac{\lambda}{\epsilon\sqrt\rho} +O(\epsilon^0) +\frac{\rho^{2+2p}}{\epsilon} +\frac{\rho}{\epsilon^2}}.
\label{223}
\end{eqnarray}
Let us choose the parameters like this: $\epsilon=\lambda^{49/100}$, $\theta=\lambda^{1/100}$, $\rho=\lambda$, $p>-1/2$. Then the minimum in \fer{223} is given by $\frac{2\theta\lambda^2}{\epsilon}\gamma$  (provided $\lambda\leq (4\gamma)^{-25/13}$) and the error term in \fer{223} is $\O{\lambda^{1/100}+O(\lambda^0)}=O(\lambda^0)$. The  virial theorem tells us that $\av{B}_{\psi_\lambda}=0$, so 
\begin{equation}
\av{P_{1,\beta} P(\Lambda\leq \lambda)}_{\psi_\lambda} \geq 1- O(\lambda^0).
\label{224}
\end{equation}
We may write \fer{224} as 
\begin{equation}
\psi_\lambda = P_{1,\beta} P(\Lambda\leq\lambda)\psi_\lambda +O(\lambda^0) =\Omega_{1,\beta}\otimes \big(P(\Lambda\leq\lambda)\chi_\lambda\big) +O(\lambda^0),
\label{225}
\end{equation}
for some vector $\chi_\lambda\in\ff\otimes\ff$ with norm $\|\chi_\lambda\|\geq 1-O(\lambda^0)$. We point out that all estimtes are uniform in $\xi$ in any compact set. This is easily seen by noticing that the only way $\xi$ enters is through the term $K_\xi$, which is uniformly bounded in $\xi$ belonging to any compact set in $\r^2$. This finishes the proof of Theorem \ref{chev}.\\

{\it Proof of Corollary 2.4.\ } We denote by $P_{1,\beta}$, $P_{\beta,0}$ and $P_{\beta,\lambda}^\xi$ the projections onto the spans of $\Omega_{1,\beta}$, $\Omega_{\beta,0}$ and $\Omega_{\beta,\lambda}^\xi$, see \fer{gibbsvector}, \fer{kmsbo} and \fer{154}. Since $\|P_{\beta,0}-P_{\beta,\lambda}^\xi\|\rightarrow 0$ as $\lambda\rightarrow 0$ (uniformly in $\xi$ in any compact set and in $\beta\geq\beta_0$, for any $\beta_0$ fixed, [FM2]) it follows that 
\begin{eqnarray*}
\psi_\lambda =(P_{\beta,\lambda}^\xi)^\perp \psi_\lambda &=& \Pbar_{\beta,0}\psi_\lambda +O(\lambda^0)\nonumber\\
&=& \left( \Pbar_{1,\beta}\otimes P_{\Omega_0}\right)\psi_\lambda +\Pbar_{\Omega_0}\psi_\lambda +O(\lambda^0)\nonumber\\
&=& \Omega_{1,\beta}\otimes \big( \Pbar_{\Omega_0} P(\Lambda\leq\lambda)\chi_\lambda\big) +O(\lambda^0)
\end{eqnarray*}
where we used \fer{seiv} in the last step. It suffices now to observe that $\Pbar_{\Omega_0} P(\Lambda\leq\lambda)$ converges strongly to zero, as $\lambda\rightarrow 0$. This follows from $\Pbar_{\Omega_0}=\Pbar_{\Omega_\ff}\otimes P_{\Omega_\ff} +\bbbone_{\ff}\otimes \Pbar_{\Omega_\ff}$, 
\begin{equation*}
P(\Lambda\leq \lambda)=\Big( P(\d\Gamma(\omega)\leq\lambda)\otimes P(\d\Gamma(\omega)\leq\lambda)\Big)P(\Lambda\leq \lambda)
\end{equation*}
and the fact that $\d\Gamma(\omega)$ has absolutely continuous spectrum covering $\r_+$ and a simple eigenvalue at zero, $\Omega_\ff$ being the eigenvector. \hfill $\blacksquare$\\

{\bf Acknowledgements.\ } I thank W. Abou Salem, J. Derezi\'nski, J. Fr\"ohlich, M. Griesemer, V. Jak\u si\'c, A. Joye, Y. Pautrat, C.-A. Pillet, L. Rey-Bellet, I.M. Sigal, S. Starr for interesting discussions. I am particularly grateful to J\"urg Fr\"ohlich for his patience in teaching me.

\end{document}